\makeatletter \def\input@path{{ACM_styles/}} \makeatother
\documentclass[sigconf]{acmart}

\AtBeginDocument{%
  }

\copyrightyear{2026}
\acmYear{2026}
\setcopyright{cc}
\setcctype{by}
\acmConference[CHI '26]{Proceedings of the 2026 CHI Conference on Human Factors in Computing Systems}{April 13--17, 2026}{Barcelona, Spain}
\acmBooktitle{Proceedings of the 2026 CHI Conference on Human Factors in Computing Systems (CHI '26), April 13--17, 2026, Barcelona, Spain}
\acmDOI{10.1145/3772318.3791437}
\acmISBN{979-8-4007-2278-3/2026/04}

\usepackage{enumitem}
\newcounter{subfig}[figure]
 
\newcommand{\makesubtag}[1]{%
  \refstepcounter{subfig}%
  \label{#1}%
}

\begin{document}

\title{Skill-Adaptive Ghost Instructors: Enhancing Retention and Reducing Over-Reliance in VR Piano Learning}


\settopmatter{authorsperrow=4}

\author{Tzu-Hsin Hsieh}
\affiliation{%
  \institution{Delft University of Technology}
  \city{Delft}
  \country{Netherlands}}
\email{thhsieh@tudelft.nl}

\author{Cassandra Visser}
\affiliation{%
  \institution{Delft University of Technology}
  \city{Delft}
  \country{Netherlands}}
\email{c.m.s.visser@outlook.com}

\author{Elmar Eisemann}
\affiliation{%
  \institution{Delft University of Technology}
  \city{Delft}
  \country{Netherlands}}
\email{e.eisemann@tudelft.nl}

\author{Ricardo Marroquim}
\affiliation{%
  \institution{Delft University of Technology}
  \city{Delft}
  \country{Netherlands}}
\email{R.Marroquim@tudelft.nl}

\renewcommand{\shortauthors}{Hsieh et al.}

\begin{abstract}
  Motor-skill learning systems in XR rely on persistent cues. However, constant cueing can induce overreliance and erode memorization and skill transfer. We introduce a skill-adaptive, dynamically transparent ghost instructor whose opacity adapts in real time to learner performance. In a first-person perspective, users observe a ghost hand executing piano fingering with either a static or a performance-adaptive transparency in a VR piano training application. We conducted a within-subjects study (N=30), where learners practiced with traditional \textit{Static} (fixed-transparency) and our proposed \textit{Dynamic} (performance-adaptive) modes and were tested without guidance immediately and after a 10-minute retention interval. Relative to \textit{Static}, the \textit{Dynamic} mode yielded higher pitch and fingering accuracy and limited error increases, with comparable timing. These findings suggest that adaptive transparency helps learners internalize fingerings more effectively, reducing dependency on external cues and improving short-term skill retention within immersive learning environments. We discuss design implications for motor-skill learning and outline directions for extending this approach to longer-term retention and more complex tasks.
\end{abstract}


\begin{teaserfigure}
  \includegraphics[width=\textwidth]{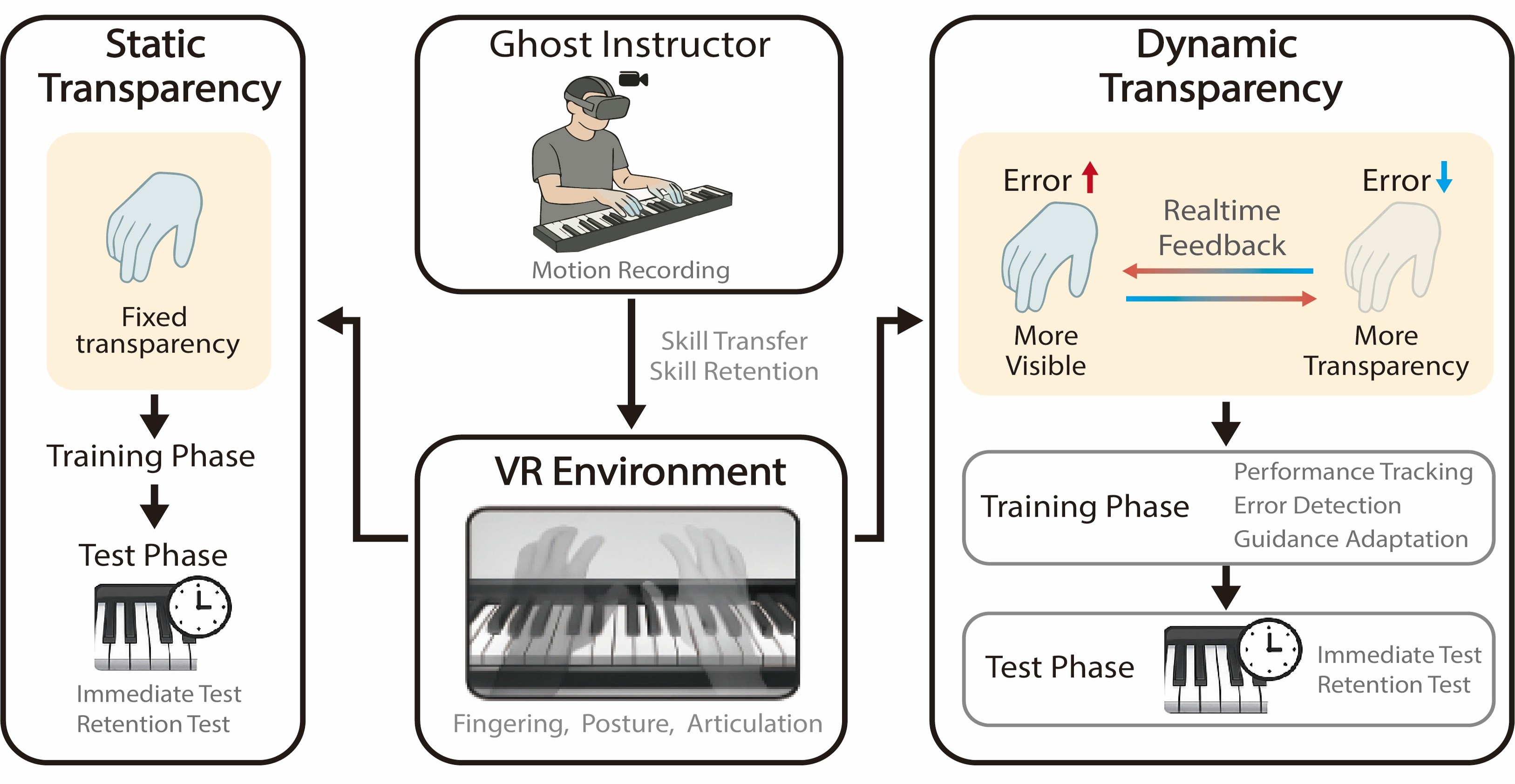}
  \Description{Overview of a VR piano training pipeline. An instructor records a reference performance that appears as a semi-transparent ghost hand; a learner practices in VR by following it. The figure contrasts a Static mode with a proposed Dynamic mode, where the learner’s pitch, timing, and fingering signals drive an adaptive controller that increases or decreases the ghost hand’s opacity during practice.}
  \caption{Skill-Adaptive Ghost Instructors: ghost-hand guidance with two modes: Static (fixed opacity) and Dynamic (opacity adapts to the learner's performance).}
  \label{fig:overview}
\end{teaserfigure}

\maketitle

\section{Introduction}

Immersive technologies such as Virtual Reality (VR) have demonstrated substantial potential for supporting \textit{motor skill learning} across diverse domains, including sports \cite{Neumann2018VRsport,DuKing2018VRinSport}, medical training \cite{seymour2002virtual}, and musical performance \cite{pesek2024enhancing, johnson2020evaluating}. These domains share common requirements: learners must coordinate spatial precision, temporal sequencing, and ergonomic control under time constraints. Piano playing is a representative case for VR-based music guidance systems, as it requires precise timing, pitch accuracy, and correct fingering. Prior research has demonstrated that VR training can support both synchronous remote instruction and in-situ assistance \cite{fuchino2023t2remoter, haar2021embodied}, enabling learners to receive real-time guidance without the physical presence of an instructor. VR environments are particularly suited for such training because they can deliver embodied demonstrations and allow learners to observe and imitate expert actions from a first-person perspective \cite{fribourg2020virtual, hulsmann2019superimposed}, thereby facilitating the acquisition and transfer of complex motor skills. 

Despite these advantages, many existing VR-based music guidance systems continue to rely heavily on persistent visual cues \cite{deja2022survey}. This often results in learners focusing on reproducing notes accurately while neglecting the development of proper fingering strategies \cite{salmoni1984knowledge, sigrist2013augmented}. Fingering accuracy is particularly important in piano performance, as it directly influences long-term skill transfer and technical fluency \cite{Candidi2014PianoTMS, engel2012learning, furuya2013transfer}.  In addition, visual guidance in these systems is generally presented at a fixed visibility level, without adapting to the learner’s performance \cite{wilson2023feedback, banquiero2023passthrough, banquiero2024color}. This lack of adaptability can lead to over-reliance on guidance \cite{sigrist2013augmented}, reduced attention to self-correction \cite{salmoni1984knowledge}, and ultimately weaker retention once the cues are removed, particularly for tasks that require both accuracy and correct technique \cite{schmidt1997continuous}. This trade-off between \textit{instructional support} and \textit{learner autonomy} represents a broader HCI challenge in the design of adaptive feedback for immersive learning.

Taken together, the evidence motivates a shift from static to adaptive visual cueing. Recent research has shown that adaptive visibility can balance instructional support and learner autonomy \cite{yang2002implementation, sigrist2013augmented, renganayagalu2021effectiveness}. While VR studies show that making avatar or object representations visible can improve motor performance and control \cite{shin2022effects, borgwardt2023vrisbee}, evidence for such visibility manipulations in fine-motor, rhythm-dependent tasks remains limited \cite{rigby2020piarno, wilson2023feedback}. Piano training provides a representative case study for this problem: learners must coordinate pitch accuracy, timing precision, and ergonomic fingering, all while maintaining rhythmic flow. These characteristics make piano an ideal domain for investigating how adaptive visual guidance can both mitigate over-reliance and support retention.

To address the limitations of persistent visual cues, we present a \textbf{skill-adaptive ghost instructor} that dynamically adjusts its transparency according to learner performance. We examine these limitations in the context of VR piano learning. The system continuously estimates pitch accuracy, timing precision, and fingering correctness, mapping them into a performance signal that drives ghost visibility in real time, reducing guidance as proficiency increases. When performance is accurate, the ghost becomes faint, encouraging learner autonomy; when errors accumulate, the ghost reappears to provide additional support. We evaluate this approach in a within-subjects user study ($N=30$) comparing two visibility modes (Static and Dynamic) across two melodies (A and B). Measured results include objective accuracy, subjective workload, and short-term retention.



Accordingly, we ask: \textit{how can adaptive transparency balance instructional support and learner autonomy in VR-based motor-skill training, and what impact does it have on retention compared with static transparency?} In summary, this paper makes three contributions to immersive motor learning:

\begin{enumerate}
  \item \textbf{Conceptual framing.} We identify over-reliance on persistent cues as a barrier to retention in VR-based motor skill training, highlighting the need for adaptive guidance that balances support with autonomy.
  \item \textbf{Design intervention.} We introduce a skill-adaptive ghost instructor that modulates its visibility in real time based on learner performance. By framing adaptive transparency as a design principle, our approach addresses first-person occlusion and fingering correctness, advancing beyond static cues and conventional fading schemes.
  \item \textbf{Empirical insights.} Through a controlled user study in VR piano training, we demonstrate that adaptive transparency improves fingering accuracy, reduces errors, lowers workload, and enhances short-term retention compared to static cues. We further distill design implications for extending adaptive visual guidance to other fine-motor training contexts such as sports practice and rehabilitation.
\end{enumerate}

\section{Related Work}
\subsection{XR-based Motor Learning Systems}

\subsubsection{General Motor Skill Training}
Extended reality (XR) technologies are increasingly used to support motor‐skill training that requires spatial precision, temporal sequencing, and ergonomic control. Critically, multiple domains report \emph{transfer} to real-world performance or \emph{error reduction}: in \emph{sports}, training has shown on-field transfer in batting \cite{gray2017transfer} and broader performance benefits across sport domains \cite{neumann2018systematic}; in \emph{surgical training}, laparoscopy simulation improves operating-room performance \cite{Seymour2002AnnSurgVR}; in \emph{clinical rehabilitation}, adding non-immersive VR to treadmill training reduced falls in high-risk older adults in a multicenter randomized controlled trial (RCT) \cite{mirelman2016addition}; and in \emph{industrial assembly}, AR guidance shortens task time and lowers error rates compared with paper/video instructions \cite{Henderson2011TVCG_MaintenanceAR,Hou2015JCCE_PipingAR,Yang2019FrontiersARAssembly}. Together, these results show that XR affords controlled, repeatable practice with task-aligned feedback and motivates skill-adaptive, fading assistance to mitigate cue dependence and promote retention across motor domains.

Observational learning is another robust pathway for acquiring motor skills: watching skilled or learning models can shape movement form, timing, and decision-making, particularly when demonstrations are structured and paired with concise cues \cite{ste2012observation, blandin_lhuisset_proteau_1999, andrieux_proteau_2016}. From a design standpoint, the constraints-led perspective emphasizes aligning practice environments with performer, task, and environmental constraints to elicit functional coordination patterns \cite{clark2019effectiveness}. Standard texts synthesize these principles into actionable guidance for feedback, practice variability, and progression. Foundational motor-learning work further motivates how feedback should be scheduled and framed: reduced or faded knowledge-of-results can yield better long-term retention than constant feedback \cite{winstein1988relative, aoyagi2019feedback}, contextual interference in practice schedules can benefit generalization \cite{boutin2010cognitive}, and an \emph{external} focus of attention reliably improves performance and learning relative to an internal focus \cite{wulf2013attentional, wulf2016optimizing}. Collectively, these findings highlight the need for XR systems that not only provide controlled, repeatable practice but also regulate feedback and adapt assistance to learner proficiency.

\subsubsection{Piano and Music Training Systems}

Within music education, piano is a representative case for XR-based guidance because it requires precise timing, pitch accuracy, and correct fingering. Prior work can be organized along two orthogonal axes: \emph{pedagogical function} and \emph{interface strategy}. On the pedagogical axis, four themes recur \cite{deja2022survey}: (i) synchronizing movement aids that help learners keep time (animated guides and mirrored or projected tutor hands \cite{xiao2014andante,xiao2010mirrorfugue,xiao2013mirrorfugue,xiao2011duet}); (ii) sight-reading support that maps notation to the keyboard (colored lines, falling-note or piano-roll cues, and on-key highlights \cite{santos2013augmented,yang2013visual,birhanu2017keynvision,cai2019design,chow2013music,rodrigues2021personalization,hackl2017holokeys,rogers2014piano,banquiero2023passthrough}); (iii) motivational designs that add curricula and gamified feedback \cite{gerry2019adept,molloy2019mixed,weing2013piano}; and (iv) improvisation aids that suggest harmonies or recognize gestures (hand tracking or surface EMG) \cite{dahlstedt2015mapping,liang2016barehanded,glickman2017music,hantrakul2014implementations}. On the interface axis, systems differ in how guidance is delivered: notation-assistance overlays project cues on or above the keys and can accelerate early learning \cite{rigby2020piarno,weing2013piano,pan2018pilot,Amm2024MRPiano}, HoloKeys superimposes target keys and supports demonstrations via AR with IoT key sensing \cite{hackl2017holokeys}, embodied demonstration/imitations foreground \emph{how} to play \cite{labrou2023following,guo2021hand} with learner-friendly viewpoints in the MirrorFugue series \cite{xiao2013mirrorfugue,xiao2010mirrorfugue,xiao2011duet}, and remote MR/VR lesson rooms provide co-presence, hand tracking, and key-press visualization for bidirectional teaching \cite{schlagowski2023wish,goagoses2024teachers,wang2020ppvr}. Taken together, these archetypes cover \emph{what} to play through cues, \emph{how} to play through embodied examples, and \emph{where} to learn through shared virtual studios.

Despite encouraging gains in note accuracy, timing, and engagement, gaps remain. Reviews of XR piano tutoring systems show that prior work largely employs \emph{fixed-visibility guidance} and emphasizes \emph{knowledge-of-results} (e.g., pitch correctness, rhythmic alignment), whereas \emph{knowledge-of-performance} (e.g., fingering, posture, articulation) is comparatively under-supported \cite{wilson2023feedback,deja2022survey,Amm2024MRPiano,sellers2024augmented}. This imbalance can foster over-reliance on visual prompts during solo practice and weaker technique formation once cues are removed \cite{mackrous2007specificity,sanford2022investigating}, which in turn has been shown to impair skill retention in piano training \cite{buchanan2012overcoming}. Explicit occlusion handling is rarely addressed: most systems maintain visibility via fixed overlays rather than dynamically managing hand–ghost–keyboard occlusion \cite{banquiero2023passthrough,wilson2023feedback}. 

These observations motivate systems that adapt both the \emph{visibility} and the \emph{content} of guidance in real time to balance support with autonomy. While AR/MR is often cited as ideal for in-situ deployment, prior research highlights VR as the preferred modality for motor-skill training that requires tight experimental control, precise feedback timing, and rigorous logging \cite{hoover2021designing, vaughan2016overview}. We leverage VR as a controllable testbed to isolate the effects of our adaptive transparency mechanism, avoiding the confounds of variable lighting or background clutter. This choice is supported by Zahabi et al. \cite{zahabi2020adaptive}, who find that adaptive training systems rely on fully virtual environments to standardize stimuli and isolate adaptive control strategies, and by Lui et al. \cite{lui2025efficacy}, who show that immersive VR is uniquely capable of supporting precise, performance-contingent adjustments during training. Consequently, our system is VR-based: we use headset passthrough for a brief calibration and alignment step, then train the target piano skills in a fully virtual scene. This setup employs embodied demonstrations with \emph{adaptive instructor visibility} to reduce cue dependence, improve transfer to the physical keyboard, and enhance retention once guidance is removed.

\subsection{Ghost Instructors and Visual Perspectives}

Perspective shapes how learners interpret demonstration visuals. A first-person perspective tends to increase embodiment and imitation for fine motor actions, while a third-person perspective reduces occlusion and supports global inspection \cite{gorisse2017first, trabucco2019user}. Comparative studies report complementary benefits for the two views, which motivate the use of perspective blending or shared viewpoints between the instructor and learner. A blended-view approach \cite{kawasaki2010collaboration} merges real-world camera feeds of both users into a single composite for head-mounted display viewing. While this reduced positional error compared to most view configurations, showing only the partner’s view produced better velocity control, suggesting that no single perspective is optimal for all performance metrics. Augmented reality telepresence has also been explored for remote guidance. The Vishnu system \cite{le2016vishnu} lets a distant expert control two virtual arms attached to the learner’s shoulders, manipulating virtual proxies aligned to real objects. This method significantly outperformed a monitor-based baseline for complex tasks, highlighting the value of spatially registered, embodied instruction without co-location.

\begin{figure*}[t]
  \centering
  \includegraphics[width=\linewidth]{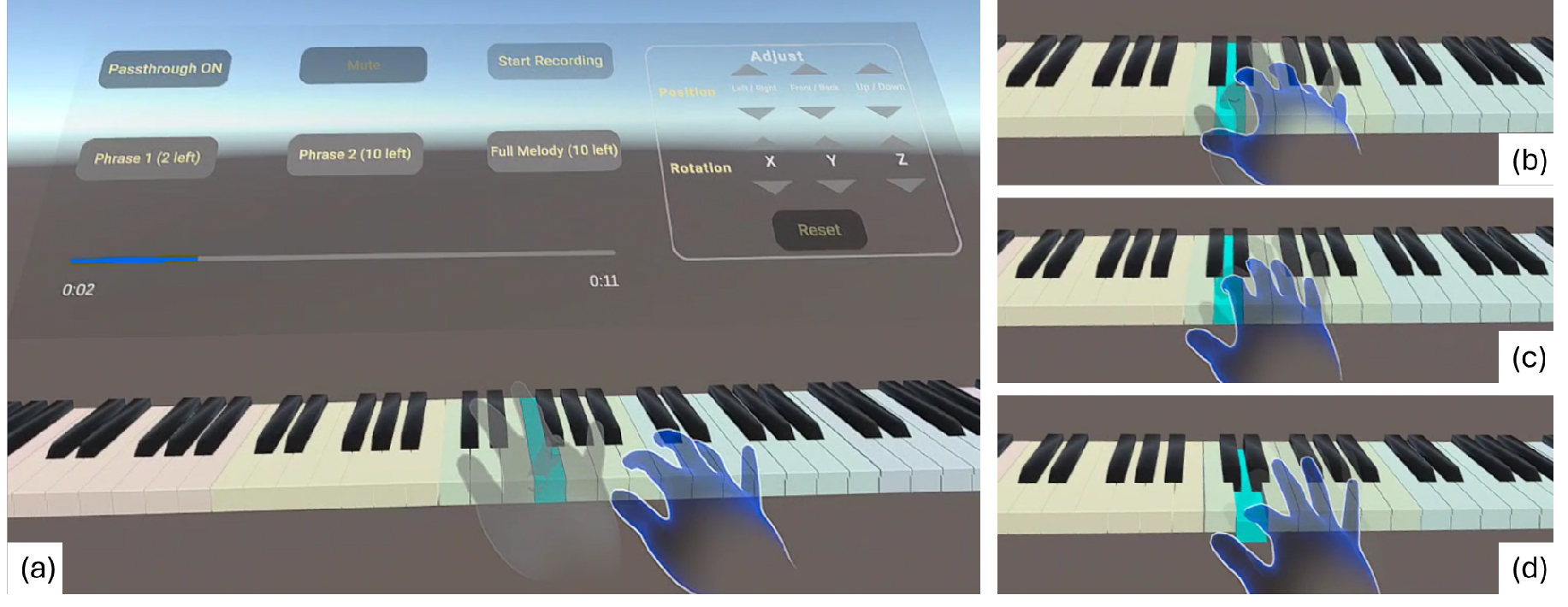}

  \Description{User interface of the VR piano trainer. Panel (a) shows controls for desk passthrough alignment, ghost audio mute/unmute, and recording; a row to select Phrase~1, Phrase~2, or Full Melody and to display remaining trials (default 10); and an Adjust panel to translate/rotate the virtual keyboard to match the physical desk. Panels (b–d) illustrate dynamic transparency: the ghost instructor hand becomes more visible when performance deteriorates and fades as pitch, timing, and fingering stabilize; the participant hand is rendered separately to reduce occlusion.}

  \caption{System interface and dynamic-transparency feedback.
  The blue hand denotes the \emph{participant}, and the gray hand denotes the \emph{ghost instructor}.
  (a) Controls. Passthrough desk alignment; ghost audio mute/unmute; record button.
  The second row selects \textit{Phrase~1}, \textit{Phrase~2}, or \textit{Full Melody} and shows the remaining-trial count (default 10).
  The \textit{Adjust} panel fine-tunes the virtual piano rig to the physical desk (translate/rotate for alignment).
  (b--d) Adaptive feedback. Ghost-hand opacity is driven by performance in real time: higher visibility under poorer performance, and fainter visibility as pitch, timing, and fingering stabilize.}

  \label{fig:ui_dynamic_feedback}
\end{figure*}

Within these perspectives, ghost-style instruction systems in VR overlay a semi-transparent, co-located tutor avatar directly onto the learner’s body to guide motion  \cite{chinthammit2014ghostman}. In Just Follow Me \cite{yang2002implementation}, the tutor’s motion was visualized as a first-person ghost aligned to the learner’s body, with swept-volume trails and acceleration vectors to convey path and dynamics. In controlled motion learning tasks, first-person, co-located guidance delivered training and transfer effects comparable to or better than indirect video imitation. Other approaches explore virtual co-embodiment \cite{fribourg2020virtual}, where teacher and student jointly control a single avatar via a weighted average of their movements. Comparative studies \cite{kodama2023effects} show that co-embodiment and first-person ghost metaphors (1PPGM) accelerate learning more than no guidance. However, both approaches suffer from performance drops when guidance is removed, with the largest drop observed in co-embodiment, a smaller drop in 1PPGM, and the smallest drop with no guidance. Beyond perspective, tuning visibility itself affects performance: in navigational obstacle-avoidance tasks, transparent or invisible hands with opaque objects yielded faster completion and fewer collisions,and are preferred by users \cite{hatira2024effect}. Notably, prior ghost-based systems fixed the tutor’s visibility, leaving the potential of adaptive transparency unexplored.



\subsection{Adaptive Support Strategies for Motor-Skill Learning}

Adaptive training refers to closed-loop personalization, continuously adjusting task challenge or scheduling feedback based on a learner's ongoing performance \citep{kelley1969adaptive,guadagnoli2004challenge}. For a system to be adaptive, three core components are required: (1) performance measurement, (2) an adaptive variable, and (3) adaptive logic \citep{zahabi2020adaptive}. Prior work on motor-skill tutoring typically adapts either \emph{how} assistance is delivered or \emph{who} provides guidance, which can be categorized into two broad streams: \emph{solo practice} and \emph{collaboration-guided} support. 

In solo practice, most systems adapt either \textit{task difficulty} or the \textit{feedback modality}. \emph{BACh} \citep{yuksel2016BACh} increases musical complexity based on cognitive workload to maintain an optimal challenge zone, whereas \emph{Adapt2Learn} \citep{turakhia2021adapt2learn} maps performance to a running score and bi-directionally raises or lowers difficulty when progress plateaus. Other work, such as \emph{Hapticus} \citep{lee2025hapticus}, explores \textit{modality selection and ordering} (EMS, linkage exoskeleton, vibrotactile gloves) through pre-specified haptic sequences that improve in-the-moment performance but do not provide closed-loop, within-practice adaptation. However, beyond hardware complexity, haptic guidance (HG) faces significant efficacy challenges. Specifically, HG can hinder learning if the guidance reduces the learner’s active involvement in error correction. Marchal-Crespo et al. \cite{marchal2013effect} demonstrated that while HG is beneficial for gross motor tasks such as tennis or rowing, visual feedback can be more effective for internalizing complex spatial topologies, as haptic assistance may induce a "passive learning state" where users fail to actively correct their own errors. Given that piano playing requires extreme fine-motor precision, complex haptic devices can often constrain natural movement and interfere with the ergonomic flow of performance. Consequently, our approach leverages dynamic transparency as a lightweight visual scaffold that guides spatial positioning without interfering with physical touch or the piano's natural tactile feedback.


Collaboration-guided approaches emphasize making demonstrations and instructions accessible and clear. \emph{Digitusync} \citep{nishida2022digitusync} mechanically couples a teacher’s and learner’s hands through a passive exoskeleton to improve imitation fidelity during co-located instruction. \emph{ShowMe} and \emph{ARTEMIS} \citep{amores2015showme,gasques2021artemis} support immersive tele-gestures and spatial annotations to strengthen shared attention in remote collaboration. These methods primarily optimize \textit{who provides guidance}, rather than algorithmically adapting support based on performance. While effective for following expert demonstrations, they can inadvertently foster over-reliance, making it unclear whether learners can perform autonomously once such support is withdrawn.

Evidence further shows that the timing and direction of adaptation critically shape learning effectiveness and efficiency \citep{guadagnoli2004challenge}, while guidance-hypothesis and faded-feedback literature warns that excessive or continuous guidance can induce dependency \citep{anderson2005support, salmoni1984knowledge, winstein1990reduced}.
Addressing this dependency requires regulating the salience of guidance over time. Sigrist et al. \cite{sigrist2015sonification} employed error-based transparency in rowing training. While effective for gross-motor trajectory learning, their visibility modulation is primarily driven by a single scalar spatial error signal, without integrating multi-dimensional performance cues. In contrast, our Ghost Instructor targets the high-dimensional requirements of piano performance, and its transparency controller integrates pitch correctness, rhythmic timing, and fingering accuracy into a unified performance signal. Moreover, we introduce an asymmetric adaptation mechanism that fades guidance out quickly as performance stabilizes while reintroducing guidance more cautiously when errors increase, aiming to encourage retrieval practice and mitigate over-reliance on continuous visual cues. 


 Building on this work, our study extends adaptive training theory in three ways. First, we introduce \emph{adaptive transparency} as a simple, purely visual adaptive variable that can be implemented on commodity XR devices, complementing hardware-intensive systems that require worn devices to adapt task difficulty or haptic actuation and may constrain users’ natural movement. Second, we instantiate a continuously corrected control strategy that maps \emph{multi-dimensional and fine-grained} piano performance signals, including not only pitch accuracy but also timing precision and finger-specific correctness, to the moment-to-moment visibility of a virtual tutor. This allows assistance to become more salient when needed and to gradually fade when the learner is performing well, while performance is continuously monitored in real time. Third, we explicitly operationalize and evaluate \emph{dependency} by comparing immediate and retention performance once guidance is removed, addressing an outcome that is rarely measured in prior motor-skill tutoring work.

\subsection{Summary}
Prior work in motor learning highlights three key requirements for XR-based training: (i) integrate observation and task-representative practice to reduce representational gaps; (ii) regulate the \emph{frequency, timing, and framing} of feedback to preserve error-driven learning while avoiding dependence; and (iii) adapt assistance dynamically to the learner’s moment-to-moment proficiency, balancing instructional support with autonomy \cite{winstein1990reducedkr, wulf2013attentional}. 

Existing XR-based piano learning systems integrate notation overlays, embodied demonstrations, and shared lesson spaces. However, they often emphasize knowledge of results more strongly than knowledge of performance, which can lead to visual over-reliance and limit skill retention. Research on instructional perspectives shows that first-person ghost overlays and co-embodiment can accelerate motor learning. However, both approaches experience performance drops when guidance is removed and have so far only employed fixed-visibility avatars.

In response to these limitations, we propose \emph{skill-adaptive ghost instructors} that modulate virtual-ghost transparency in real time based on the learner's performance in note accuracy, timing precision, and fingering correctness. Within adaptive training theory, our design instantiates a different adaptive variable: rather than changing task difficulty or switching feedback modalities, we regulate the \emph{visual availability} of a tutor. Conceptually, we treat ghost visibility as a limited guidance resource whose salience must be matched to the learner’s current proficiency to avoid reliance on assistance. By tailoring visibility to moment-to-moment performance while holding the task constant, our approach aims to reduce assistance dependence, maintain engagement, and improve retention within a VR piano setting.

\section{Research Question and Hypotheses}\label{sec:rq-hypotheses}


\noindent\textbf{RQ1.} What effects does performance-adaptive transparency, compared with static transparency, have on objective learning outcomes (transfer and short-term retention) in VR-based motor-skill training?

\noindent\textbf{RQ2.} How does performance-adaptive transparency, compared with static transparency, shape the balance between instructional support and learner autonomy?

\subsection{Hypotheses}\label{sec:hypotheses}

Guided by motor-learning theory on guidance dependence and feedback scheduling, we hypothesized the following effects of \textit{Dynamic} (performance-adaptive transparency) versus \textit{Static} (fixed transparency):

\textbf{H1 (Skill transfer).}\label{hyp:h1} Dynamic transparency will promote better transfer when switching to a new melody and condition. Participants who trained under Dynamic first will start a new melody and condition at a higher initial performance than those who trained under Static first.  

\textbf{H2 (Retention).}\label{hyp:h2} Dynamic transparency will improve short-term retention relative to static transparency. The decrease in the composite performance score from the Immediate Test to the Retention Test will be smaller for Dynamic than for Static.

\textbf{H3 (Training convergence).}\label{hyp:h3} Across training, both conditions will show positive practice effects, but Dynamic transparency will lead to faster convergence than Static. Participants in the Dynamic condition will show larger gains in early loops and smaller marginal improvements in later loops, or will reach a predefined performance criterion in fewer loops, compared with Static.


\section{Methodology}

\subsection{System Overview}
Our VR piano training system (Figure \ref{fig:overview}) is designed to improve fingering, posture, and articulation, with the goal of enhancing both skill transfer and short-term skill retention. The setup uses a ghost instructor that appears as an overlaid virtual hand performing the target fingering. Two guidance modes are supported: a static transparency mode with fixed visibility, and a performance-adaptive mode where the ghost’s transparency changes in real time according to the learner’s fingering accuracy and timing performance.

Expert demonstrations are recorded using hand and finger tracking to capture joint positions and movements in 3D space, along with key press events and audio. These recordings are then replayed in a fully virtual piano scene aligned to the learner’s real keyboard. During practice, the ghost instructor provides real-time, moment-by-moment guidance. The adaptive transparency further encourages learners to focus on self-correction when performance is accurate. After training, users perform an Immediate Test and a Retention Test without any ghost guidance to evaluate skill transfer and retention.

\subsection{Motion and Input Capture}
\subsubsection{Hand and Finger Tracking}\label{sec:hand_finger_tracking}
We employ the Meta XR Hands API for 26 joint skeletal tracking of the learner’s right hand in real time. Joint positions and rotations are streamed at the device’s native update rate in the headset’s local coordinate system, and then transformed into the calibrated piano coordinate frame. This alignment allows us to directly compare the learner’s finger positions with the target ghost-hand motion.

The tracking data serves two main purposes: (1) detecting fingering correctness in real time, and (2) providing the basis for performance-aware transparency adaptation. Fingering correctness is evaluated by mapping tracked fingertip joints to pressed keys and comparing them with the intended fingering from the ghost demonstration. In addition, per-frame hand-posture data (joint positions and rotations) are recorded for subsequent analysis in Section~\ref{subsec:hand_ghost_similarity}. Timing precision and note accuracy are computed from synchronized key event logs, enabling integrated assessment of pitch, rhythm, and fingering to drive adaptive feedback.

\subsubsection{Motion Recording}
\label{sec:motion_recording}
Building on the tracking setup, we log the 26-joint skeleton at a fixed 30fps with per-frame timestamps, storing joint rotations and positions in the calibrated piano frame. For animation export, positions are recorded internally. However, emitting position curves is optional, allowing a trade-off between playback fidelity and file size.

For each included joint, we store a timestamped local pose (relative to the hand root) and later serialize the sequence into a Unity \texttt{AnimationClip}. Keyframe curves are compressed using rotation and position slope thresholds to reduce file size without losing perceptible motion fidelity.

The motion recording module serves two purposes in our system:
\begin{enumerate}[topsep=3pt,itemsep=3pt,parsep=3pt]
    \item \textbf{Ghost Instructor Demonstrations} – We first record our own expert performances, producing smooth, reusable ghost-hand animations that users follow during the training phase.
    \item \textbf{User Performance Capture} – During the test phase, users’ hand motions are recorded in the same format. These recordings are not used in real-time feedback but are stored for \textit{post-hoc} analysis of timing, fingering correctness, and articulation patterns.
\end{enumerate}

This dual-use design standardizes instructor and learner motion data, enabling accurate playback, direct comparison, and kinematic analysis. In this proof-of-concept, instructor motions were author-recorded and kept identical across all participants and conditions. Therefore, any stylistic bias applies equally across participants and conditions, and cannot be used to explain condition effects. Future work will include multiple expert performers.

\subsubsection{Audio and Key Event Logging}
\label{sec:audio_key}
We developed a logger to capture every press–release cycle on the virtual keyboard. For each key event, the system records the key identifier (pitch), onset time, press duration, and finger index. A dictionary of active keys tracks onset times until release, when the duration is computed and the event is stored. Recording can be toggled on/off.

Two modes match the study workflow: demonstration recordings capture the author’s ghost–hand performances, while user recordings capture participants’ play during the test phase. All sessions are saved as JSON files with versioned names, time–aligned with motion recordings to support post-hoc analysis of note accuracy, timing precision, sustain characteristics, and fingering correctness.

\subsection{Playback and Visualization}

The playback system replays pre-recorded ghost-hand motions in the VR environment to guide learners during practice. Learners view the scene from a first-person perspective to promote embodiment and imitation. The ghost hand is displayed as a semi-transparent grey model, overlaid on the virtual piano and spatially aligned so that each animated press matches the learner’s corresponding virtual key. To clearly differentiate between the instructor and the learner, the participant’s own tracked hands are rendered in semi-transparent blue. Additionally, a key highlight system provides note-specific visual cues by changing the color of pressed keys to light blue; thus indicating both the precise timing and location of each note.

To structure practice, the interface provides three playback modes: \textit{Phrase~1}, \textit{Phrase~2}, and \textit{Full~Melody}, where \textit{Phrase~1} and \textit{Phrase~2} together constitute the \textit{Full~Melody}. Each mode replays its corresponding ghost-hand performance ten times consecutively, with a two-second countdown shown between repetitions to allow participants to prepare. This segmentation encourages focused practice on smaller sections before attempting the full piece, while ensuring consistent repetition and exposure to each segment.

\subsection{Adaptive Transparency Control}
\label{sec:adaptive_transparency}
To mitigate over-reliance on visual guidance while still supporting learners who struggle, we implemented an adaptive transparency controller that dynamically adjusts the ghost hand’s visibility based on real-time performance metrics. For each frame during playback, the system evaluates the learner’s input against the reference melody and ghost avatar motion, producing normalized sub-scores for pitch correctness, timing precision, and fingering correctness, denoted $s_p$, $s_t$, and $s_f$, respectively, with $s\in[0,1]$.

The overall performance score is a weighted sum
\begin{equation}
S = w_p s_p + w_t s_t + w_f s_f,
\label{eq:score}
\end{equation}
\noindent\textbf{Calculating Sub-scores.}
These sub-scores are calculated in real time based on the user's input. \textbf{Pitch correctness ($s_p$)} assigns full credit when the pressed key's pitch matches the target and the onset falls within a temporal window $\pm\tau$ of the reference time; otherwise $s_p=0$. In other words, a note is treated as correct only if both pitch and timing are within this acceptance window.
\textbf{Timing precision ($s_t$)} is computed as $1 - \min (\frac{\left| \Delta t \right|} {\tau}, 1)$ where $\Delta t$ is the onset deviation (ms) relative to the reference time and $\tau = 120$\,ms (approximately 0.20--0.25 beats at 100--120\,BPM), a musically meaningful window that also accommodates VR latency.
\textbf{Fingering correctness ($s_f$)} uses semantic finger labels (Thumb–Index–Middle–Ring–Pinky) from the hand-tracking API. We apply a two-level partial-credit rule: $s_f = 1$ for the intended finger, $s_f = 0.5$ if the used finger is adjacent to the intended one on the same hand, and $s_f = 0$ otherwise. To robustly handle individual differences in hand size, comparisons rely on these semantic labels rather than absolute 3D joint coordinates.


The corresponding error that drives transparency is
\begin{equation}
E = 1 - S,
\label{eq:error}
\end{equation}
with $S,E\in[0,1]$. Here, $E$ denotes the instantaneous error signal (higher means worse performance and thus higher target opacity). 

Our adaptive transparency score is a weighted sum of pitch, timing, and fingering correctness. Prior work indicates that pitch, timing, and fingering are co-learned and mutually reinforcing in beginner training. Within this interplay, curricula and assessments often place early emphasis on pitch accuracy, while allowing some timing flexibility and introducing fingering strategies progressively  \cite{Goebl2013,DallaBella2011,Allingham2022,Parncutt1997}. Based on this rationale, we adopted the configuration $w_p=0.7$, $w_t=0.2$, $w_f=0.1$, which prioritizes note accuracy while still incorporating rhythm stability and ergonomic fingering. While this weighting is grounded in prior pedagogical research, the optimal balance between pitch, timing, and fingering may vary across learners and tasks.

\noindent\textbf{From error to opacity.}
We first stabilize frame-to-frame fluctuations with an asymmetric exponential moving average (EMA) on the error:
\begin{equation}
\hat{E}_t=\lambda_t E_t+(1-\lambda_t)\hat{E}_{t-1},\qquad
\lambda_t=\begin{cases}
\lambda_{\uparrow}, & \text{if } E_t>\hat{E}_{t-1}\\
\lambda_{\downarrow}, & \text{otherwise}
\end{cases}
\label{eq:ema_asym}
\end{equation}
$E_t$ is the per-frame value of $E$, and $\hat{E}_t$ is its temporally smoothed version. Here $\lambda_t\in(0,1)$ is the smoothing factor: larger values make $\hat{E}_t$ follow $E_t$ more quickly, while smaller values yield heavier smoothing. We choose a smaller rise rate $\lambda_{\uparrow}$ and a larger decay rate $\lambda_{\downarrow}$ so that the ghost fades quickly when the learner improves and reappears more cautiously after occasional slips. This reduces cue dependence and promotes attention to pitch accuracy, rhythmic alignment, and proper fingering. Importantly, this EMA introduces only brief within-stream smoothing ($\approx$0.5–1\,s) to stabilize visuals. This short latency operates within a single performance stream and is therefore distinct from the multi-trial feedback delays (terminal feedback) typically associated with enhanced retention in motor-learning studies.

The filtered error is then mapped linearly to opacity and clamped to the display range:
\begin{align}
\tilde{\alpha}_t &= \alpha_{\min}+(\alpha_{\max}-\alpha_{\min})\,\hat{E}_t, \label{eq:alpha_raw}\\
\alpha_t &= \min\{\max(\tilde{\alpha}_t,\alpha_{\min}),\,\alpha_{\max}\}. \label{eq:alpha_clamp}
\end{align}

To ensure legible yet unobtrusive guidance, we clamp opacity to $\alpha_{\min}=0.08$ and $\alpha_{\max}=0.8$. Thus, a low filtered error $\hat{E}_t$ yields a faint ghost that does not occlude the keyboard, whereas a high error makes the ghost clearly visible without becoming fully opaque. We also apply a small per-frame visual easing when updating the material to avoid shader jitter; this does not alter the linear mapping in Eqs.~\ref{eq:alpha_raw}–\ref{eq:alpha_clamp}. In our study, we compare a \textit{Static} condition (constant $\alpha=0.5$) with a \textit{Dynamic} condition (adaptive opacity as described above).

\subsection{Implementation Details}
We developed the system in Unity 2022.3.44f1 (LTS) targeting Meta Quest 3. For rapid iteration, the headset was tethered to the development PC via Oculus Link over USB-C, enabling in-headset Play-in-Editor without producing Android builds during prototyping. The project uses the Meta XR All-in-One SDK (v69.0.1, October 2024). From this SDK, we employed the camera-rig prefab for HMD tracking, the interaction building blocks (e.g., buttons and UI interactors) for interface elements, and the hand-tracking modules to capture and drive hand-based interactions. This configuration provided a stable, up-to-date integration path for Quest features while keeping iteration times low.

\section{User Studies}

\subsection{Experiment Design}\label{sec:design}
We employed a within-subject design with two experimental conditions:
\begin{itemize}
    \item \textbf{Static Ghost Hand}: a semi-transparent ghost hand with fixed visibility.
    \item \textbf{Dynamic Ghost Hand}: a ghost hand whose transparency is adapted in real time according to the learner’s performance (pitch accuracy, timing precision, fingering correctness).
\end{itemize}

Each participant practiced and performed two short piano melodies under both conditions. 
The order of conditions (Static first vs.\ Dynamic first) and melody assignment (Melody A vs.\ Melody B) were counterbalanced across participants using a Latin square design, resulting in four groups (SA/DB, SB/DA, DA/SB, DB/SA) (see Figure~\ref{fig:latin_square_balance}). 
Each session began with a short familiarization phase, followed by practice and performance trials.

After completing both conditions, participants filled out the NASA-TLX workload questionnaire and additional Likert-scale questions. 
Objective performance metrics (note accuracy, timing deviations, and fingering correctness) were logged automatically during the trials.

Design Rationale. We acknowledge that the Dynamic condition represents a composite intervention consisting of performance-adaptive transparency and a sub-second stabilization latency ($\approx$ 0.5–1s) introduced by the asymmetric EMA. This delay was intentionally incorporated to prevent visual jitter and to implement a "cautious reappearance" strategy: when errors occur, the instructor ghost returns gradually rather than abruptly, encouraging the learner to attempt self-correction before relying on the visual prompt. Consequently, the observed learning benefits in the Dynamic condition likely stem from the synergistic effect of both transparency adaptation and this brief feedback delay.

\subsection{Participants}

A total of \textbf{30} participants (\textbf{18 male, 12 female}; age \textbf{22--39} years, $M=27.6$, $SD=4.33$) took part in the study. 
A power analysis ($\alpha=.05$, two-sided) indicated that with $N=30$ our within-subjects design has  approximately 80\% power to detect medium-to-large effects ($d_z \approx 0.55$; $f \approx 0.50$, $\eta^2_p \approx .20$). 
This sample size provides adequate power for medium-to-large effects; detecting smaller effects would require a larger sample. 
Half of the participants ($n=15$) reported having prior piano experience (self-rated score $\geq 3$ on a 5-point scale), while the other half ($n=15$) had little or no experience ($<3$). 
All participants had prior experience with Virtual Reality (VR) and were comfortable using VR headsets (self-rated $\geq 3$). 
All had normal or corrected-to-normal vision. The study was approved by an institutional review board committee.

\begin{figure}[tb]
  \centering 
  \includegraphics[width=\columnwidth]{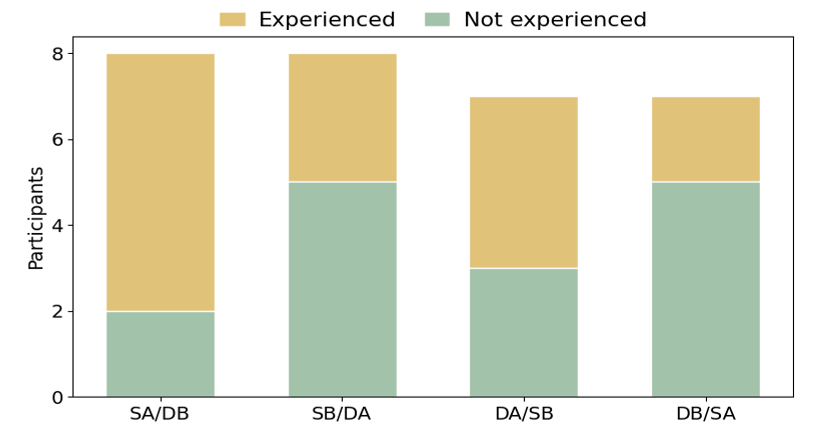}
  \Description{Stacked bar chart showing participant counts per Latin-square order, split by piano experience (experienced vs not experienced). Orders are SA/DB, SB/DA, DA/SB, and DB/SA. Totals are 8, 8, 7, and 7 participants respectively. By experience: SA/DB has 6 experienced and 2 not experienced; SB/DA has 3 experienced and 5 not; DA/SB has 4 experienced and 3 not; DB/SA has 2 experienced and 5 not.}
  \caption{Participant distribution across Latin-square conditions and piano experience. S = \textit{Static}, D = \textit{Dynamic}; each condition pairs a visibility mode with a melody A/B.}
  \label{fig:latin_square_balance}
\end{figure}

Participants were assigned to one of four counterbalanced groups following a Latin square design (Figure~\ref{fig:latin_square_balance}). Each group included both experienced and inexperienced pianists, though the distribution was not perfectly even (e.g., SA/DB: 2 inexperienced, 6 experienced; SB/DA: 5 inexperienced, 3 experienced). A chi-square test confirmed that the proportion of experienced vs.\ inexperienced participants did not differ significantly across groups ($\chi^2(3, N=30) = 3.93, p = .27$).

\subsection{Task Representation}

Participants were asked to learn and perform two short piano melodies in VR. Each melody was approximately 12 seconds in length, contained \emph{single notes only}, and was performed with the \emph{right hand only}. During practice, the system rendered a ghost hand co-located with the participant’s virtual hand to illustrate intended finger placement and motion. 

Each session began with a brief familiarization, followed by two phases: (1) a \emph{training} phase, during which participants followed the ghost hand to rehearse the melody, and (2) a \emph{test} phase, in which they were asked to play as accurately and steadily as possible without the ghost hand. The system automatically logged key events, per-finger joint trajectories, loop indices, and timing information. During training, the system computed the per-loop, per-note sub-scores (Eq.~\ref{eq:score}) and the aggregated error (Eq.~\ref{eq:error}) in real time to drive ghost transparency, and logged these values for offline analysis.


\subsection{Procedure}

Each session lasted approximately 50–60 minutes per participant (pre-study ~5 min; two blocks × [training ~10 min + immediate test (2 trials) + 10-min retention interval + retention test (2 trials) + questionnaires ~5 min]).

\paragraph{Pre-study}
After providing informed consent, participants completed a brief pre-study questionnaire assessing piano experience and VR/AR experience and comfort, followed by headset fitting and a short calibration/familiarization to align the virtual keyboard slightly above the physical desk height, ensuring keys could be played on the tabletop rather than in midair.

\paragraph{Experimental blocks}
Each participant then completed two counterbalanced blocks, one per condition (Static and Dynamic). In each block, the training phase comprised three progressive parts with the condition’s ghost hand visualization active: Phrase 1 (first subphrase; 10 loops), Phrase 2 (second subphrase; 10 loops), and Full Melody (Phrase 1 + Phrase 2; 10 loops). The training lasted about 10 minutes. Immediately after training, an Immediate Test was administered consisting of two trials; the better trial was retained for analysis. About 10 minutes later, participants completed a Retention Test, again consisting of two trials, where the better trial was retained. During the interval, participants solved an expert-level Sudoku as a distractor to discourage rehearsal or phone use. Following the retention test of each block, participants completed the NASA TLX assessment as well as the post-condition Likert items. A short break was offered before switching to the other block.

\paragraph{Post-study}
After the completion of both blocks, participants filled out a post-study survey capturing overall preference (Static vs.\ Dynamic), comparative ratings, and open-ended feedback.

\subsection{Measures}
We collected both objective performance measures and subjective responses.

\paragraph{Performance Measures.}  
We use the same pitch, timing, and fingering definitions as in Section~\ref{sec:adaptive_transparency}.
Briefly, pitch accuracy counts a note as correct only if the pressed key pitch matches the target and the onset falls within the temporal window $\pm\tau$; timing precision is computed as $1 - \min(|\Delta t|/\tau, 1)$; and fingering correctness follows the partial-credit rule over semantic finger labels (intended vs.\ adjacent vs.\ other fingers).
During the Training Phase, we compute these sub-scores per loop to analyze learning curves; during the Test Phase, we aggregate them per trial for inferential statistics. The error rate \(E\) is the signal that drives adaptive transparency and is defined as \(E=1-S\), where \(S=w_p s_p + w_t s_t + w_f s_f\) (weights as in Section~\ref{sec:adaptive_transparency}). In the \textit{Training Phase}, we logged pitch accuracy, finger accuracy, timing accuracy, and error rate for each practice loop to analyze the learning curve.  
In the \textit{Test Phase}, we recorded (a) pitch, timing, and fingering via the audio and key event logging system (Section~\ref{sec:audio_key}) to compute accuracy and error measures, and (b) 3D hand motion trajectories via motion capture (Section~\ref{sec:motion_recording}) for exploratory motion similarity analysis. Unless otherwise specified, all error bars in the figures indicate standard error of the mean (SEM).  

\paragraph{Subjective Measures.}  
Participants rated statements on 5-point Likert scales (1 = strongly disagree, 5 = strongly agree). During the pre-study, we collected demographics and musical background. Post-study measures included: \textit{Learning Experience} (7 items), \textit{Visual Design \& Transparency} (4), and \textit{Retention} (7). We also included a \textit{Comparative evaluation} with two Likert items (“easy to learn from” for \textit{Static} and \textit{Dynamic}) and three forced-choice questions (Static/Dynamic/Not sure) on overall preference, memorization support, and naturalness. Perceived workload was assessed with the \textit{NASA--TLX}.







\begin{figure*}[!t]
  \centering
  \includegraphics[width=\textwidth]{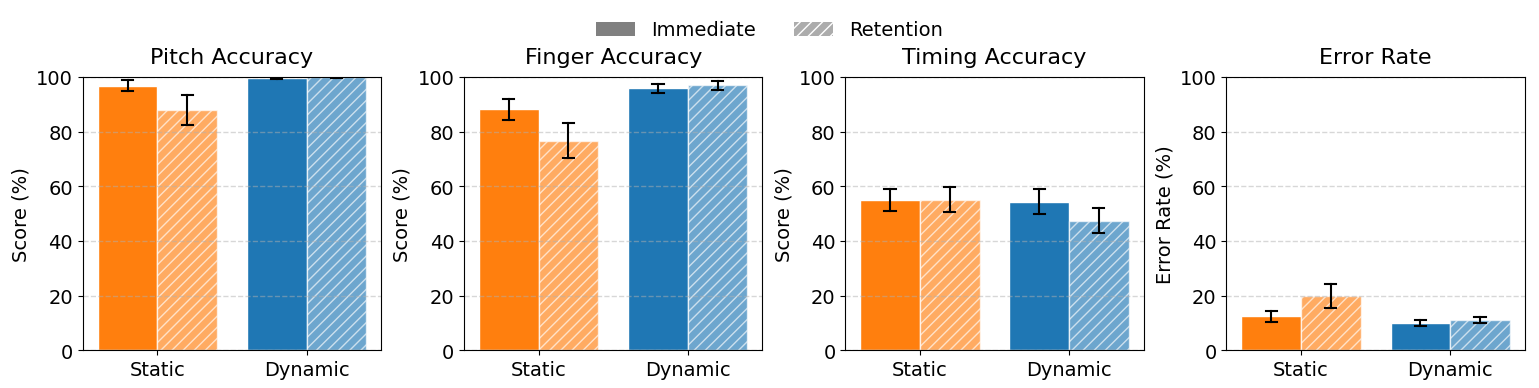}

  \Description{Four-panel bar charts comparing Static and Dynamic guidance in Immediate and Retention tests. Pitch and Finger accuracy are higher under Dynamic than Static, with smaller Immediate-to-Retention declines for Dynamic. Timing accuracy appears similar across conditions. Error rate is lower for Dynamic, with a smaller increase at Retention. Error bars indicate SEM.}
  
  \caption{ Performance comparison between Static and Dynamic ghost hand conditions across pitch accuracy, finger accuracy, timing accuracy, and error rate. Error bars represent the standard error of the mean (SEM).}
 \label{fig:dynamic_static}
\end{figure*}

\section{Results}
\label{sec:results}

\subsection{Performance Results}
This section presents objective performance results. Analyses of Static vs.\ Dynamic, effects of prior experience, and Immediate vs.\ Retention use the participants’ \emph{performance test data} (Immediate/Retention), while learning-curve progression is based on \emph{training phase} practice logs.

\subsubsection{Static vs. Dynamic}
As shown in Figure~\ref{fig:dynamic_static}, we compared participants’ performance between the \textit{Static ghost hand} and \textit{Dynamic ghost hand} conditions across four objective measures: pitch accuracy, finger accuracy, timing accuracy, and error rate. A two-way repeated-measures ANOVA with factors \textit{Condition} (Static vs. Dynamic) and \textit{Test Type} (Immediate vs. Retention) was conducted for each measure.  

\paragraph{Pitch Accuracy.} 
There was a significant main effect of Condition ($F$(1,29) = 5.04, $p$ = 0.03), 
with higher pitch accuracy in the Dynamic condition compared to the Static. 
As shown in Table~\ref{tab:dynamic_static}, participants achieved nearly ceiling-level accuracy 
in the Dynamic condition (99.1\% in the Immediate test, 99.6\% in the Retention test), 
whereas accuracy dropped under the Static condition, particularly in the Retention test (96.8\% Immediate vs. 87.8\% Retention).  
This indicates that the advantage of Dynamic guidance was consistent across both Immediate and Retention tests.    

\paragraph{Finger Accuracy.}  
A strong main effect of Condition was found ($F$(1,29) = 9.68, $p$ < 0.01), 
showing that Dynamic guidance significantly outperformed Static. 
Numerically, finger accuracy was stable and high in the Dynamic condition (95.8–96.5\%), 
but decreased in the Static condition from 88.2\% in the Immediate test to 76.7\% in the Retention test (Table~\ref{tab:dynamic_static}).

\paragraph{Timing Accuracy.}  
No significant effects were observed for Condition ($F$(1,29) = 1.42, $p$ = 0.24), 
Test Type ($F$(1,29) = 0.88, $p$ = 0.36), or their interaction ($F$(1,29) = 1.04, $p$ = 0.32). 
As shown in Table~\ref{tab:dynamic_static}, timing performance remained comparable across conditions 
and test phases, ranging between 46.8\% and 55.0\%.

\paragraph{Error Rate.}  
A significant main effect of Condition was found ($F$(1,29) = 4.29, $p$ = .047), with Dynamic yielding fewer errors. 
Specifically, error rates in the Dynamic condition remained low (9.9\% Immediate, 11.6\% Retention), 
while errors increased more strongly in the Static condition, particularly in the Retention test (12.4\% Immediate vs. 19.8\% Retention; Table~\ref{tab:dynamic_static}).  

\begin{table}[H]
    \centering
    \small
    \Description{Accessible summary. The table lists mean percentages for four measures—PitchAcc, FingerAcc, TimingAcc, and ErrorRate—by condition (Static, Dynamic) and test (Immediate, Retention). Higher is better for PitchAcc, FingerAcc, and TimingAcc; lower is better for ErrorRate. Dynamic is higher in PitchAcc and FingerAcc in both tests (Immediate: 99.1\% vs 96.8\% and 95.8\% vs 88.2\%; Retention: 99.6\% vs 87.8\% and 96.5\% vs 76.7\%). ErrorRate is lower with Dynamic (Immediate: 9.9\% vs 12.4\%; Retention: 11.6\% vs 19.8\%). TimingAcc is similar at Immediate (~55\%) but lower with Dynamic at Retention (46.8\% vs 55.0\%).}
    
    \caption{Mean performance scores (in \%) for Static vs. Dynamic ghost hand conditions in Immediate and Retention tests.}
    \label{tab:dynamic_static}
    \resizebox{\linewidth}{!}{%
    \begin{tabular}{l l c c c c}
        \toprule
        Condition & Test Type & PitchAcc & FingerAcc & TimingAcc & ErrorRate \\
        \midrule
        Static  & Immediate & 96.79 & 88.16 & 55.00 & 12.43 \\
                & Retention & 87.84 & 76.71 & 55.04 & 19.83 \\
        Dynamic & Immediate & 99.11 & 95.84 & 54.36 & 9.85 \\
                & Retention & 99.56 & 96.52 & 46.81 & 11.61 \\
        \bottomrule
    \end{tabular}}
\end{table}

\paragraph{Summary.}  
Across performance measures, the Dynamic ghost hand consistently outperformed the Static condition. Specifically, Dynamic guidance significantly improved pitch accuracy and finger accuracy, while also reducing overall error rates. Timing accuracy did not statistically differ between conditions. As reported above for timing, the Condition $\times$ Test Type interaction was not significant, indicating that the Dynamic–Static difference did not vary between Immediate and Retention.

\subsubsection{Immediate Test vs. Retention Test}

\begin{figure}[t]
  \centering
  \includegraphics[width=\linewidth]{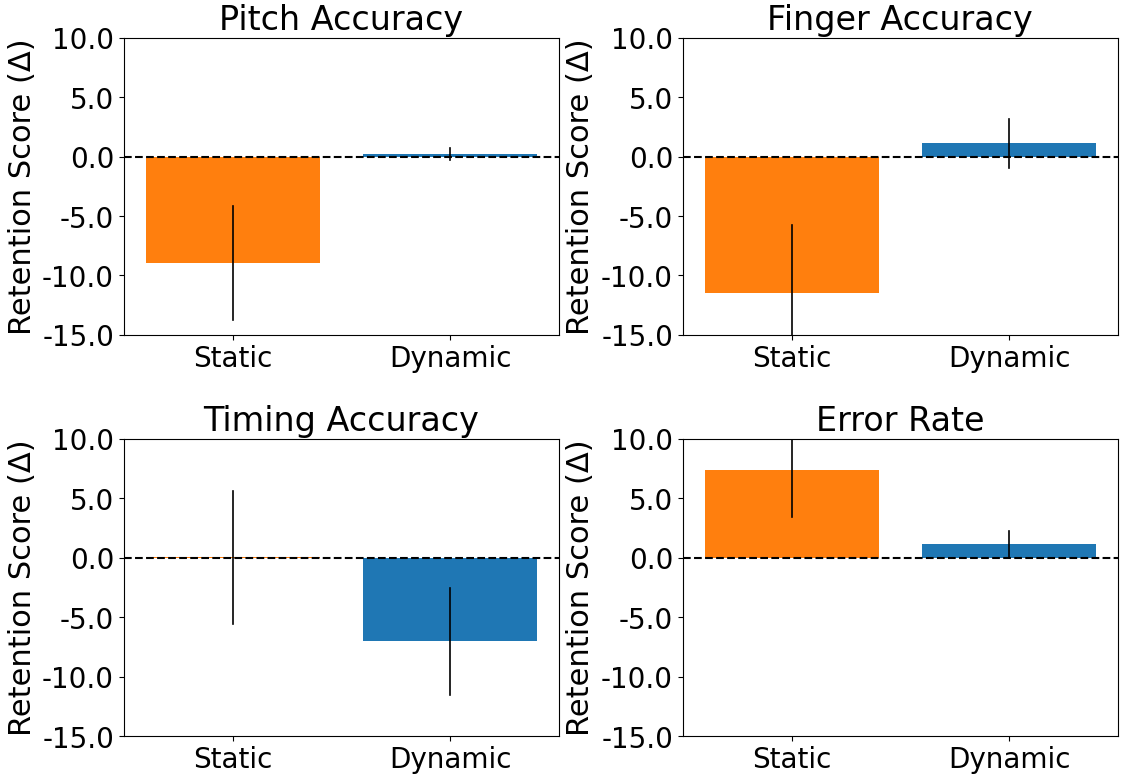}
  \Description{
  Four bar plots compare retention scores (Retention – Immediate) between Static (orange) and Dynamic (blue) ghost hand conditions. 
Pitch and finger accuracy decline under Static but remain stable or improve under Dynamic. 
Timing accuracy stays near zero for Static but decreases under Dynamic. 
Error rate rises strongly in Static while remaining low in Dynamic.
  }
  \caption{
  Retention scores (Retention -- Immediate) for pitch, finger, and timing accuracy as well as error rate, comparing Dynamic and Static ghost hand conditions. Values near zero indicate stable performance, with negative values showing decline and positive values showing improvement.
  }
  \label{fig:retention_score}
\end{figure}

To assess memory for the learned material, we computed a \textit{retention score} for each metric as the difference between the Retention and Immediate tests (Retention$-$Immediate). Scores near zero indicate stable performance, negative scores indicate decline, and positive scores indicate improvement. We compared retention scores between the Static and Dynamic conditions using paired $t$-tests for each metric. As shown in Figure~\ref{fig:retention_score}, retention patterns differed by condition.

For \textit{pitch accuracy}, Static showed a marked decline ($\Delta=-9.0 \pm 4.8\%$), whereas Dynamic was essentially preserved ($\Delta=+0.2 \pm 0.5\%$). \textit{Finger accuracy} followed the same pattern: Static dropped substantially ($\Delta=-11.4 \pm 5.7\%$), while Dynamic remained close to baseline ($\Delta=+1.1 \pm 2.1\%$). \textit{Timing accuracy} revealed the opposite tendency, with Static showing no reliable change ($\Delta=0.0 \pm 5.6\%$) and Dynamic decreasing ($\Delta=-7.0 \pm 4.5\%$). Finally, \textit{error rate} increased in both conditions, but the rise was larger for Static ($\Delta=+7.4 \pm 4.0$ percentage points) than for Dynamic ($\Delta=+1.1 \pm 1.1$ points).

Overall, Dynamic ghost hand guidance not only yielded higher immediate performance but also supported better retention for pitch and fingering accuracy, and limited the growth of errors over time, whereas Static guidance was associated with larger declines and a greater increase in errors.

\begin{figure}[t]
    \centering
    \includegraphics[width=\linewidth]{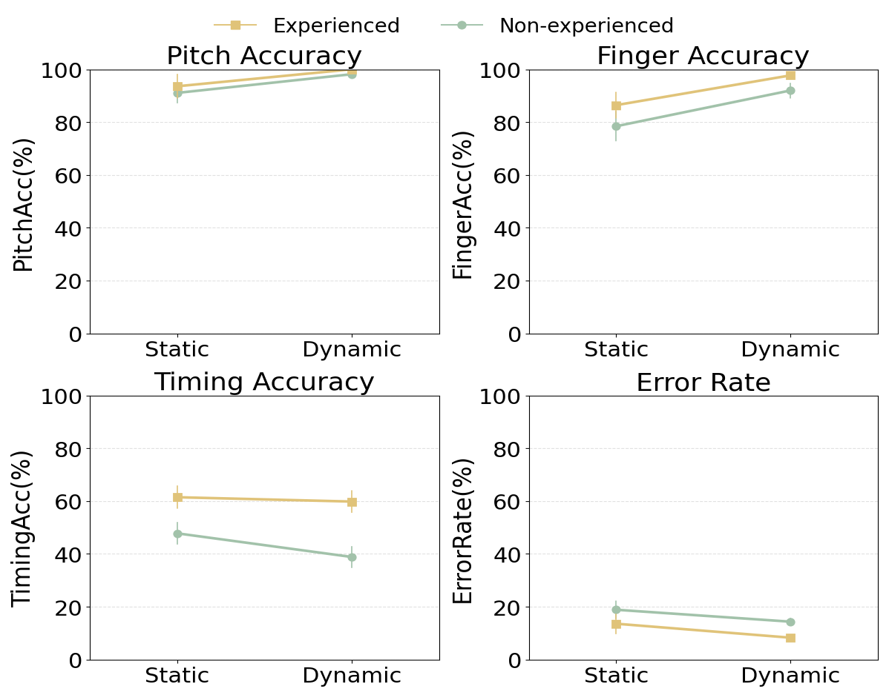}
    \Description{
    Four line plots compare performance between Static and Dynamic conditions for experienced and non-experienced participants. 
    Pitch and finger accuracy are higher for experienced participants, with both groups improving under Dynamic. 
    Timing accuracy stays stable for experienced participants but declines for non-experienced under Dynamic. 
    Error rates are consistently lower for experienced participants, and Dynamic reduces errors in both groups.
    }
    \caption{Interaction plots showing performance across Static and Dynamic conditions for experienced and non-experienced participants.}
    \label{fig:condition_experience}
\end{figure}

\subsubsection{Effect of Experience}

We tested whether prior piano experience moderated performance using a two-way design with factors \textit{Condition} (Static vs.\ Dynamic) and \textit{Experience} (experienced vs.\ non-experienced). As shown in Figure~\ref{fig:condition_experience}, both groups follow the same pattern across metrics: Dynamic yields higher pitch and finger accuracy and lower error rates than Static, whereas timing differences are small for experienced learners but more variable for inexperienced learners. Experienced participants are slightly better overall, and the lines are roughly parallel across metrics except for timing accuracy. This may reflect that experienced musicians often have a stronger sense of rhythm. 

Across these measures, ANOVAs confirmed a main effect of Condition (Dynamic~$>$~Static; e.g., pitch $F(1,116)=4.74$, $p=.032$) and no reliable main effect of Experience nor Condition$\times$Experience interaction (all $p \ge .12$). Thus, the benefit of Dynamic guidance is robust regardless of prior musical background.

\begin{figure*}[!t]
 \centering
 \Description{
 Four line plots show group learning curves over 10 practice loops for Static (orange) and Dynamic (blue) conditions. 
 Pitch and finger scores steadily improve across loops, with Dynamic slightly higher in pitch. 
 Timing scores remain relatively stable with little separation between conditions. 
 Error scores decline across practice, with Dynamic generally yielding lower  errors.
 }
 \includegraphics[width=\textwidth]{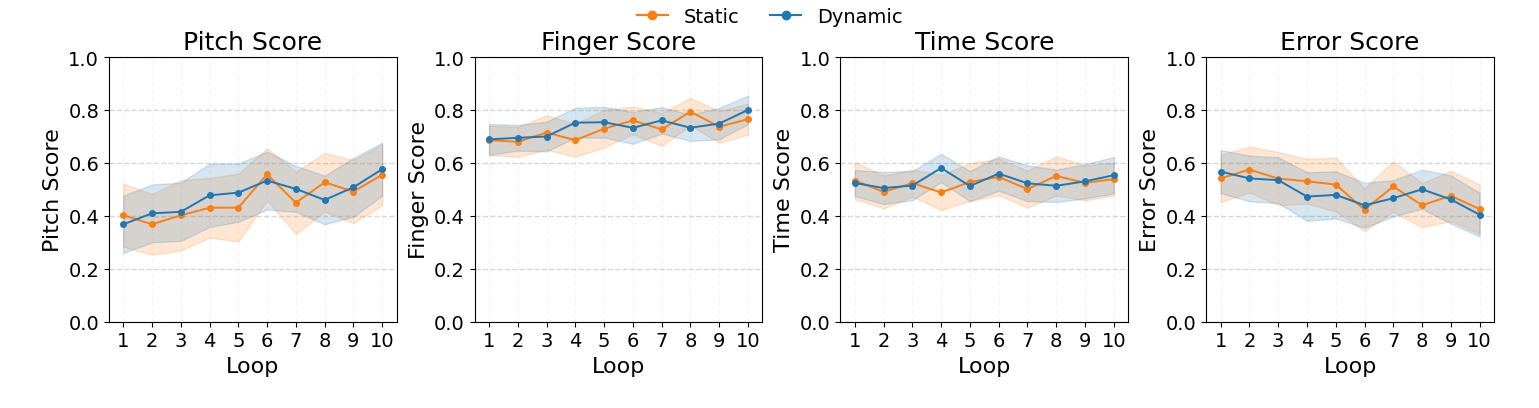}
 \caption{
Group learning curves for Full Melody: Pitch, Finger, Time, and Error.
}
 \label{fig:training_curve}
\end{figure*}

\subsubsection{Learning curve}
\label{subsec:learning-curve}

We modeled performance across ten practice loops in the training phase with a linear mixed-effects model (LMM) including fixed effects of \textit{Condition} (Static vs.\ Dynamic), \textit{Loop} (mean-centered), and their interaction, while controlling for \textit{Melody} (A/B) and \textit{Phrase} (Phrase1, Phrase2, Full Melody). Participant-specific random intercepts and random slopes for \textit{Loop} were included.

\paragraph*{Group learning curves.}
Figure~\ref{fig:training_curve} shows group-averaged learning curves for the \textit{Full Melody} across \textit{Pitch}, \textit{Finger}, \textit{Timing}, and \textit{Error} under \textit{Static} and \textit{Dynamic} guidance (lines = participant means; shaded bands = 95\% CIs).  
During practice, the two modalities follow essentially the same trajectory: \textit{Pitch} and \textit{Finger} increase, \textit{Error} decreases, and \textit{Timing} remains approximately stable. A mixed-effects model (\emph{score} $\sim$ \emph{condition} * \emph{loop\_c} + \emph{melody} + \emph{phase} + $(1{+}\,\emph{loop\_c}\mid\emph{user})$, REML) confirms improvement across loops ($\beta=0.014$, $SE=0.003$, $z=4.64$, $p<.001$), comparable learning rates (Condition$\times$Loop: $\beta=-0.003$, $SE=0.004$, $z=-0.90$, $p=.368$), and only a small average \textit{Dynamic} offset ($\beta=0.028$, $SE=0.011$, $z=2.57$, $p=.010$). Melody~B exceeds A, and later phases are associated with slightly lower learning scores Phrase~1 (Phrase~2: \(\beta=-0.044,\,SE=0.013,\,p=.001\); Full Melody: \(\beta=-0.053,\,SE=0.013,\,p<.001\)).

\begin{figure}[t]
  \centering
  \includegraphics[width=\linewidth,trim=0 0 0 0,clip]{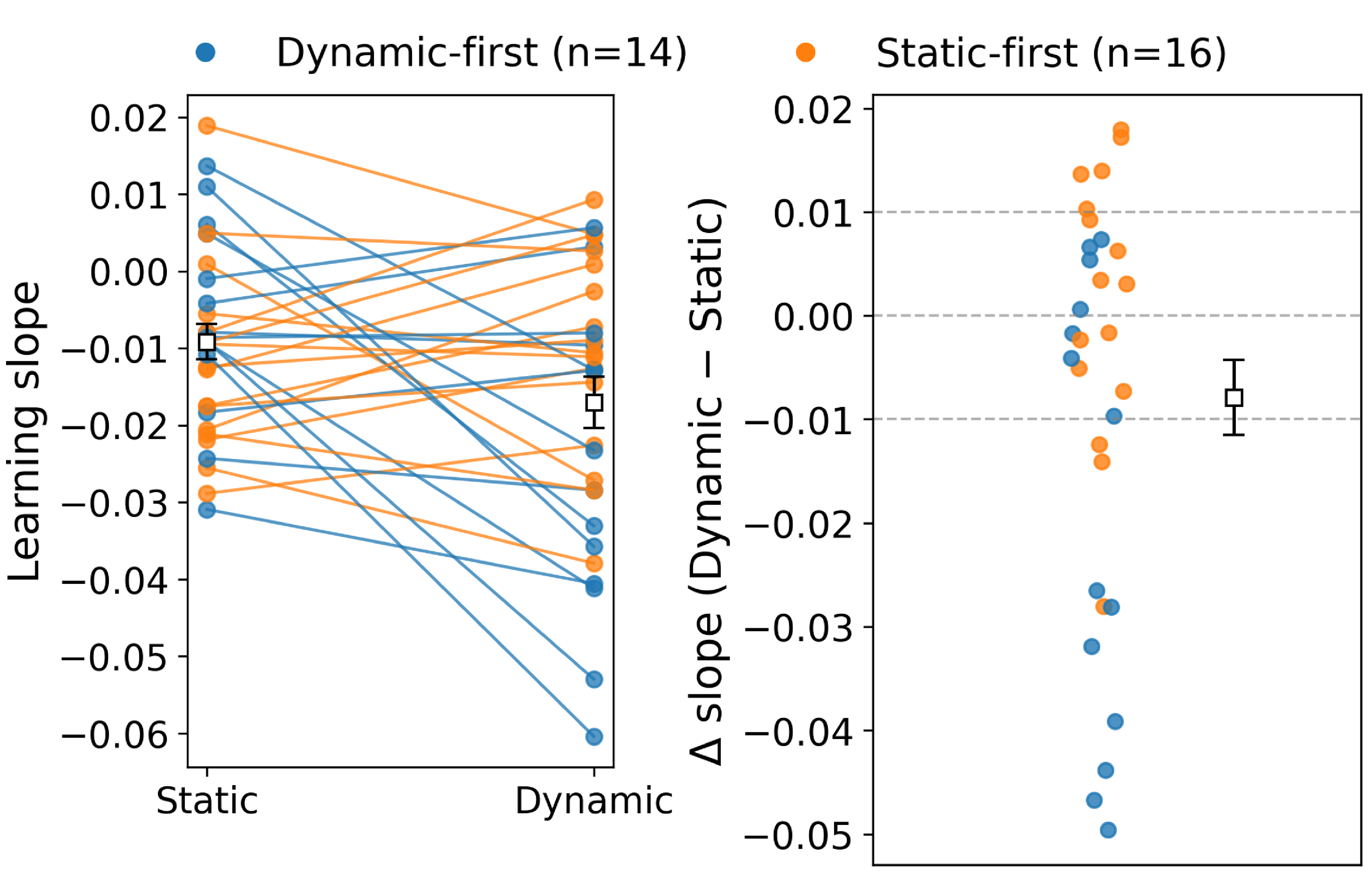}

  \begin{minipage}[t]{0.49\linewidth}
    \centering\footnotesize (a) Individual learning slopes
    \makesubtag{fig:indiv_errorrate:a}
  \end{minipage}\hfill
  \begin{minipage}[t]{0.49\linewidth}
    \centering\footnotesize (b) Per-participant difference
    \makesubtag{fig:indiv_errorrate:b}
  \end{minipage}

  \Description{Two panels summarize per-participant learning slopes for error rate (change across practice loops; more negative indicates faster improvement). Panel (a) shows paired slopes for Static and Dynamic for each participant with group means and 95\% confidence intervals; group means are slightly more negative under Dynamic. Panel (b) plots within-subject differences (Dynamic minus Static) by training order (Dynamic-first n=14, Static-first n=16); most values are below zero, indicating steeper error-rate reduction with Dynamic across orders, with individual variability present.}
  \caption{Per-participant \emph{error-rate} learning slopes by training order. white squares = mean \(\pm\)95\% CI. Lower (more negative) is better. 
  (a) Individual learning slopes for Static vs.\ Dynamic; lines link the same participant; 
  (b) Within-subject differences \(\Delta=\)Dynamic\(-\)Static grouped by order.}
  \label{fig:indiv_errorrate}
\end{figure}

\paragraph*{Individual differences.}
Figure~\ref{fig:indiv_errorrate}(a) shows per-participant \emph{error-rate} learning slopes (\textbf{lower is better}), color-coded by training order (blue = \textit{Dynamic-first}, orange = \textit{Static-first}; lines connect within subject; white squares = mean \(\pm\)95\% CI). More negative values indicate faster error reduction. Slopes were negative in both conditions (Static: \(-0.009\), \(SE=0.002\), 95\% CI \([-0.0138,-0.0049]\); Dynamic: \(-0.017\), \(SE=0.003\), 95\% CI \([-0.024,-0.010]\)), indicating decreasing error over practice, with a steeper (more negative) decline under \textit{Dynamic}.
Figure~\ref{fig:indiv_errorrate}(b) shows within-subject differences \(\Delta=\)Dynamic\(-\)Static. The mean \(\Delta\) was negative and its 95\% CI excluded zero (mean \(=-0.0077\), \(SE=0.0036\), 95\% CI \([-0.015,-0.0003]\), \(n=30\)); Cohen’s \(d_z\approx0.40\), \(p=.038\)). Critically, the effect depended on order: \textit{Dynamic-first} (\(n=14\)) showed a clear negative shift (mean \(\Delta=-0.019\), 95\% CI \([-0.031,-0.006]\), \(p=.006\)), whereas \textit{Static-first} (\(n=16\)) centered near zero (mean \(\Delta=+0.002\), 95\% CI \([-0.005,0.008]\), \(p=.642\)). The order groups differed (Welch \(t=3.08\), \(p=.006\)), indicating that adaptive transparency accelerates error reduction primarily when experienced first.

\begin{figure}[t]
  \centering
  \includegraphics[width=\linewidth,trim=0 0 0 0,clip]{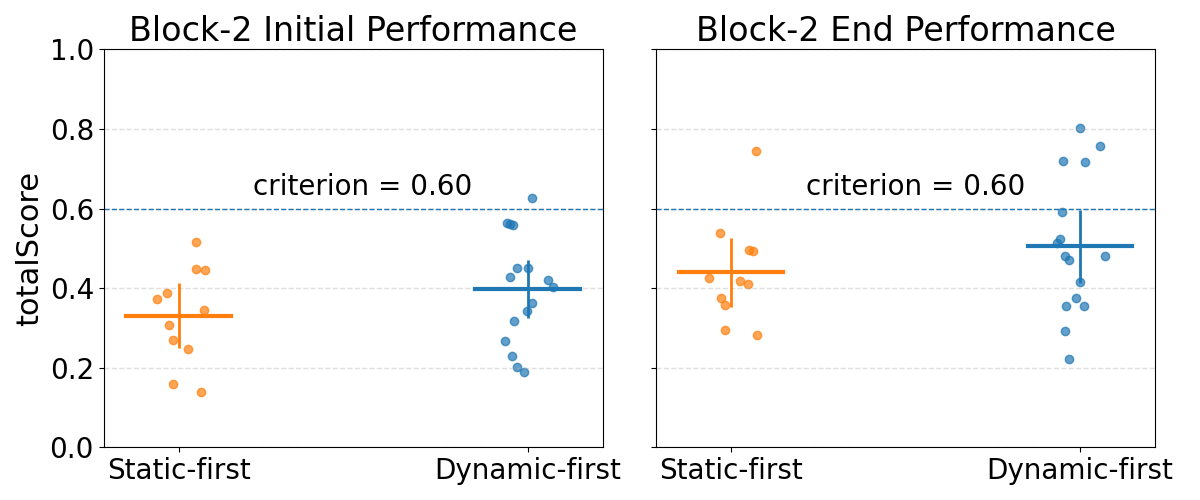}

  \begin{minipage}[t]{0.49\linewidth}
    \centering\footnotesize (a) Second-block start: initial performance (mean of the first two loops)
  \end{minipage}\hfill
  \begin{minipage}[t]{0.49\linewidth}
    \centering\footnotesize (b) End of the second block: performance (mean of the last two loops)
  \end{minipage}

  \Description{Two-panel figure showing Block-2 performance (totalScore, 0–1; higher is better) by training order. Panel (a) reports initial performance at the start of Block 2 (mean of the first two loops); Panel (b) reports end performance (mean of the last two loops). Each dot is a participant; crosshairs mark group means; a dashed line indicates the 0.60 performance criterion. At the start of Block 2, the Dynamic-first group has a higher mean than the Static-first group, with both below 0.60. By the end of Block 2, both groups increase and Dynamic-first remains higher; several Dynamic-first participants exceed 0.60, though group means remain near or below the criterion. Orange = Static-first; blue = Dynamic-first.}

  \caption{Summarizes the block switch from the first to the second block. Dots represent participants; crosshairs indicate group means; and the dashed line marks the 0.60 criterion.
  Orange = \textit{Static}-first, blue = \textit{Dynamic}-first.}
  \label{fig:switch_initial_and_end}
\end{figure}

Figure~\ref{fig:switch_initial_and_end} summarizes the block switch from the first to the second block. Figure~\ref{fig:switch_initial_and_end}(a) indicates that, at the switch to the second block (new melody and condition), participants who trained with \textit{Dynamic} first began the new melody and condition with higher initial performance, operationalized as the mean of the first two loops, than those who trained with \textit{Static} first. Using a linear mixed-effects model that includes melody (A/B) as a covariate, and a participant random intercept \( (\emph{InitialScore} \sim \emph{ConditionFirst} + \emph{Melody} + (1 \mid \emph{Participant}) )\)
, we estimate an adjusted Dynamic-first advantage of $\beta=+0.073$ ($p=.050$). Melodies differed in difficulty, with higher scores on melody~B ($\beta=+0.123$, $p=.006$). Critically, the Dynamic-first advantage holds after controlling for melody and is also visible within each melody (melody~A: $0.336$ vs.\ $0.266$; melody~B: $0.461$ vs.\ $0.385$ for Dynamic-first vs.\ Static-first). By the end of the second block, computed as the mean of the last two loops, the \textit{Dynamic}-first group still performs better, with more individuals exceeding the 0.60 criterion, whereas \textit{Static}-first rarely does (Figure~\ref{fig:switch_initial_and_end}(b)). Taken together, these distributions indicate a transfer advantage at the switch that persists over subsequent practice, while also revealing substantial between-participant variability in how quickly learners improve within the second block.

\begin{figure}[t]
  \centering
  \includegraphics[width=\linewidth]{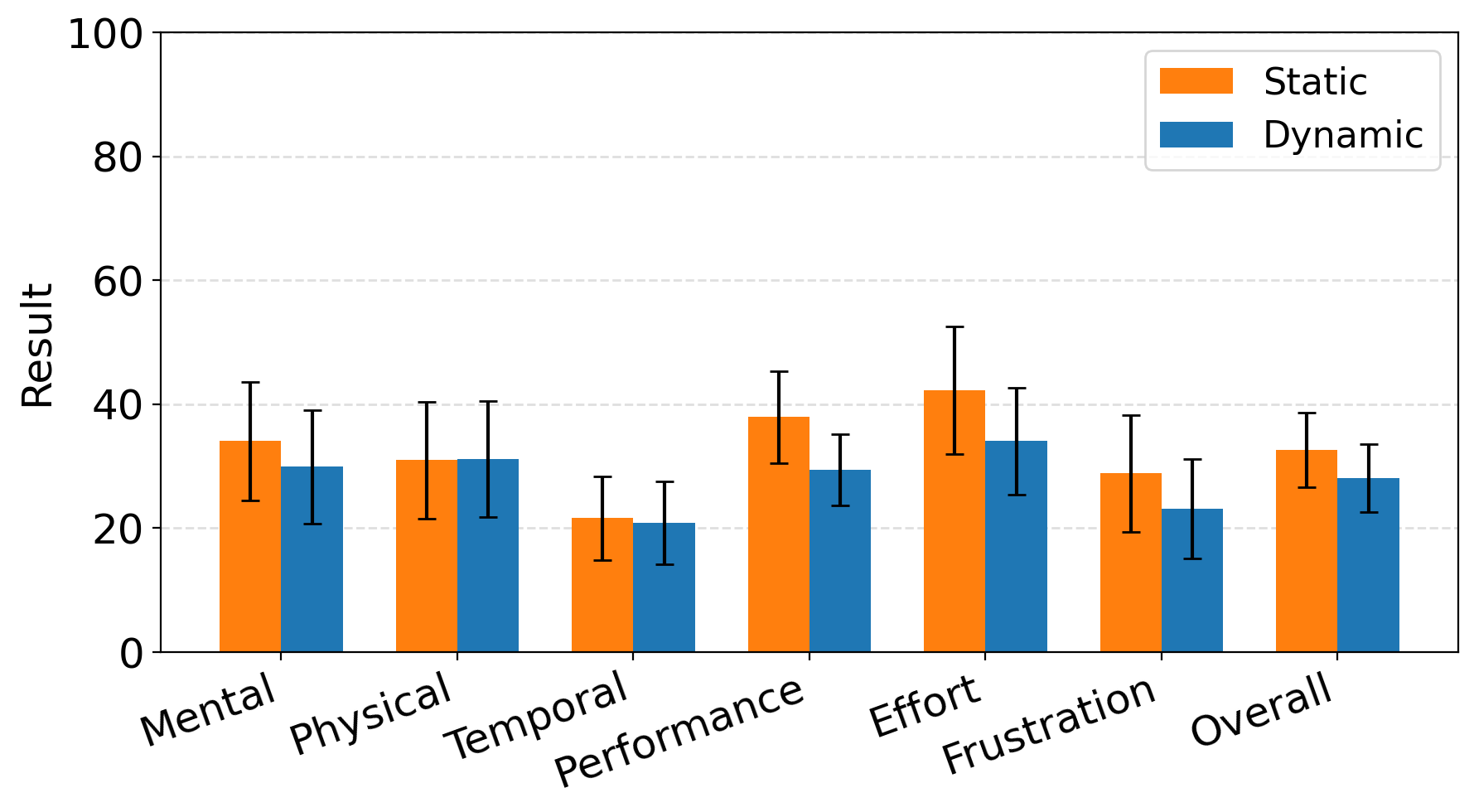}
  \Description{
  Bar chart of NASA–TLX workload subscales comparing Static (orange) and Dynamic (blue) conditions. 
  Across mental, physical, temporal, performance, effort, frustration, and overall workload, Dynamic tends to yield slightly lower scores than Static, indicating reduced workload, though differences are modest.
  }
  \caption{The NASA–TLX results for the  \textit{Static} and \textit{Dynamic} conditions (lower indicates lower workload).}
  \label{fig:nasatlx}
\end{figure}

\subsection{Subjective Results}

We measured participants’ subjective workload with NASA–TLX after each condition. Post-study, participants provided learning-experience ratings for \textit{Static}/\textit{Dynamic}, completed a Visual Design \& Transparency survey for the \textit{Dynamic} ghost, and a comparative evaluation. All items used a 5-point Likert scale.

\subsubsection{NASA-TLX}
We compared perceived workload between \textit{Static} and \textit{Dynamic} conditions with the NASA Task Load Index (NASA--TLX), which comprises six dimensions: \emph{Mental demand}, \emph{Physical demand}, \emph{Temporal demand}, \emph{Performance}, \emph{Effort}, and \emph{Frustration} (see Figure~\ref{fig:nasatlx}). Lower scores indicate less workload. 
Overall workload was lower with \textit{Dynamic} than with \textit{Static}  ($\Delta_{\text{Dyn--Sta}}=-4.5$ on the 0--100 scale, 95\% CI [$-5.9$, $-3.2$]; Holm-adjusted $p<.001$; large within-subject effect $d_z=1.30$). 


Across the individual NASA--TLX dimensions, \textit{Dynamic} reduced perceived workload relative to \textit{Static}. The largest reductions were in \emph{Performance} ($\Delta_{\text{Dyn--Sta}}=-8.5$) and \emph{Effort} ($\Delta_{\text{Dyn--Sta}}=-8.2$), followed by \emph{Frustration} ($\Delta_{\text{Dyn--Sta}}=-5.8$) and \emph{Mental demand} ($\Delta_{\text{Dyn--Sta}}=-4.2$).  
These effects remained significant after Holm adjustment ($p\le .013$) and were medium to large in magnitude ($d_z=0.57$--$1.16$). In contrast, differences in \emph{Temporal demand} ($\Delta_{\text{Dyn--Sta}}=-0.7$) and \emph{Physical demand} ($\Delta_{\text{Dyn--Sta}}=+0.1$) were small and not reliable ($p\ge .14$).


Together, adaptive transparency primarily reduced perceived performance difficulty/success cost, effort, and frustration, with smaller effects on mental demand and no change in physical demand.

\subsubsection{Post-study Questionnaires}

\textbf{Reliability.} All questionnaires used 5-point Likert scales. Internal consistency was acceptable for Learning Experience in \textit{Static} ($\alpha=.646$) and good in \textit{Dynamic} ($\alpha=.885$); the Visual Design \& Transparency scale (Dynamic-only) was also good ($\alpha=.805$). Retention reliabilities were lower (\textit{Static} $\alpha=.627$, \textit{Dynamic} $\alpha=.587$), so retention phase estimates should be interpreted with caution.

\textbf{Learning experience and Retention.}
Participants rated \textit{Dynamic} higher than \textit{Static} during practice ($t(29)=3.834$, $p=.0006$, $d_z=0.700$; $\Delta=0.529$, 95\% CI $[0.258,\,0.799]$) and during the retention phase ($t(29)=4.317$, $p=.0002$, $d_z=0.788$; $\Delta=0.452$, 95\% CI $[0.247,\,0.658]$).

\textbf{Comparative evaluation and Preference.}
The comparative Likert items were numerically higher for \textit{Dynamic} (3.867 vs.\ 3.533; $t(29)=1.980$, $p=.057$, $d_z=0.361$; $\Delta=0.334$). Preference responses strongly favored \textit{Dynamic}: across three items ($30\times3=90$ responses), 37 were marked ``Not sure'' and excluded, leaving $N=53$ decisive responses; of these, $47$ (\textbf{88.7\%}) preferred \textit{Dynamic} (Wilson 95\% CI $[0.774,\,0.947]$; binomial $z=5.632$, $p<.001$). Facet-level results were consistent with the overall finding: Dynamic exceeded Static on Overall, Memorization, and Naturalness.

\begin{table}[t]
\centering
\footnotesize
\setlength{\tabcolsep}{3pt}
\Description{Per-participant mean hand–ghost motion similarity, normalized 0–1 (higher indicates greater similarity). Columns report per-user means (μ) for Static and Dynamic in Immediate and Retention tests. Overall means show Dynamic > Static at both times: Immediate 0.610 vs 0.465 (p=0.000743, Cohen’s d_z=0.688); Retention 0.611 vs 0.414 (p=0.000021, d_z=0.925). Individual values vary, but the group pattern favors Dynamic at both time points.}

\caption{Hand similarity per user. Static/Dynamic $\times$ Immediate/Retention.}
\label{tab:hand_similarity}
\resizebox{\columnwidth}{!}{%
\begin{tabular}{l|r|r|r|r}
\hline
\textbf{User Index} & \textbf{Static (Imm.) $\mu$} & \textbf{Dynamic (Imm.) $\mu$} & \textbf{Static (Ret.) $\mu$} & \textbf{Dynamic (Ret.) $\mu$} \\
\hline
1  & 0.641 & 0.775 & 0.18 & 0.523 \\
2  & 0.349 & 0.652 & 0.315 & 0.584 \\
3  & 0.689 & 0.741 & 0.569 & 0.934 \\
$\vdots$ & $\vdots$ & $\vdots$ & $\vdots$ & $\vdots$ \\
29 & 0.558 & 0.657 & 0.486 & 0.703 \\
30 & 0.590 & 0.800 & 0.549 & 0.862 \\
\hline
\textbf{Overall} & \textbf{0.465} & \textbf{0.610} & \textbf{0.414} & \textbf{0.611} \\
\textbf{t-Test}  &             & $p=0.000743$ &             & $p=0.000021$ \\
\textbf{Cohen's $d_z$} &       & $0.688$ &            & $0.925$ \\
\hline
\end{tabular}%
}
\end{table}

\subsection{Exploratory Analysis}

\subsubsection{Hand--Ghost Motion Similarity}\label{subsec:hand_ghost_similarity}

We aligned users’ right-hand joint trajectories to the ghost reference via key-event–based linear time fitting and resampled them onto the reference timeline. Comparisons used the user–ghost joint overlap in wrist-anchored coordinates: (i) position cosine; (ii) velocity cosine on derivatives; and (iii) wrist→joint direction (frame-wise cosine; median angular error). As an alignment-robust check, DTW mean step cost is computed on position, direction, or velocity signals. Condition effects (Dynamic vs.\ Static) were tested within subjects (Immediate/Retention) using paired $t$-tests with Cohen’s $d_z$; per-user means ($\mu$) are summarized in Table \ref{tab:hand_similarity}.

In a within-subjects comparison (N=30), the \textit{Dynamic} transparency mode outperformed \textit{Static} both immediately and at retention. 
Immediate test: $M_{\text{Dyn}}=0.610$ vs.\ $M_{\text{Stat}}=0.465$, $\Delta=0.144$, $95\%\,\text{CI}=[0.069,\,0.216]$, paired $t$-test $p=0.000743$, Cohen's $d_z=0.688$. 
Retention test: $M_{\text{Dyn}}=0.611$ vs.\ $M_{\text{Stat}}=0.414$, $\Delta=0.197$, $95\%\,\text{CI}=[0.124,\,0.272]$, $p=0.000021$, $d_z=0.925$. 
Effects are medium to large for the retention test, indicating better short-term learning retention under \textit{Dynamic}.

\section{Discussion}
\subsection{Quantitative Findings}

Our central finding is that \textit{Dynamic} guidance yields significantly better retention than \textit{Static}, with higher pitch accuracy, higher fingering accuracy, and lower error rates—directly addressing over-reliance on persistent cues in VR motor-skill training \cite{sigrist2013augmented}. These outcomes align with \textbf{H2 (Retention)} and provide clear empirical support for the \textit{guidance hypothesis}, which holds that excessive or continuous feedback, characteristic of our \textit{Static} condition, suppresses error processing during practice and degrades later performance \cite{salmoni1984knowledge, schmidt1997continuous}. Prior ghost/co-embodiment studies have similarly noted that fixed-visibility guidance can accelerate early learning yet incurs a performance drop once guidance is removed \cite{kodama2023effects, yang2002implementation}. By strategically fading visibility during proficient passages and restoring it after slips, our \textit{Dynamic} controller mitigates this drop, yielding stable retention across pitch, fingering, and overall errors. Furthermore, surveys of XR piano systems \cite{deja2022survey, wilson2023feedback} highlight that prior work predominantly uses \emph{fixed}-visibility cues (e.g., \cite{rigby2020piarno, Amm2024MRPiano}). We directly address this gap by adapting \emph{knowledge-of-performance} cues—specifically, fingering visibility—and demonstrate significant gains over \textit{Static} on retention-oriented outcomes.

Unlike prior adaptive-support approaches that primarily vary \emph{how} it is delivered (e.g., haptic modality \cite{lee2025hapticus} or task-difficulty actuation \cite{yuksel2016BACh, turakhia2021adapt2learn}), or \emph{who} provides guidance (e.g., remote or dyadic co-embodiment \cite{nishida2022digitusync}), our method varies \emph{how much} guidance is provided at any moment. Treating tutor visibility as a continuously allocatable resource allows us to scale support in real time, preserving learner autonomy during practice while still enabling expert demonstration when needed. In doing so, we directly mitigate the over-reliance that co-embodiment systems frequently encounter.

Our system contributes a simpler, and purely visual method that achieves the same conceptual goal: balancing instructional support and learner autonomy, as conceptualized by Guadagnoli et al. in the Challenge Point framework \cite{guadagnoli2004challenge}. The success of our approach is further reinforced by the subjective perception results. Participants perceived the \textit{Dynamic} condition as having significantly lower workload, particularly in dimensions of perceived performance, effort, and frustration. This convergence of improved subjective experience and objective retention (H2) indicates that adapting visibility is a highly effective method for regulating instructional support. During the training phase, while group-average learning rates were similar, the per-participant slope analysis revealed that \textit{Dynamic} guidance produced significantly steeper error-reduction slopes. This supports \textbf{H3 (Training convergence)}. Moreover, learners who started with \textit{Dynamic} began the second block (new melody) at a higher initial performance level, supporting \textbf{H1 (Skill transfer)}. This suggests that the benefits of faded feedback are not limited to retention, but also promote the generalization of self-correction skills, a known failing of continuous guidance which can degrade transfer \cite{schmidt1997continuous}.

By contrast, we found no reliable timing advantage. This boundary condition suggests that visibility adaptation alone does not provide sufficient temporal scaffolding. Two plausible interpretations are that the controller emphasized correctness cues more than temporal structuring, and that the interface lacked explicit temporal cues (e.g., a beat indicator or metronome). It is also possible that learners relied on their own internal timing once note targets became familiar, reducing the marginal benefit of visibility adaptation for rhythm. Consistent with the specificity-of-practice principle \cite{mackrous2007specificity}, our controller prioritized pitch ($w_p{=}0.7$) over timing ($w_t{=}0.2$), and qualitative feedback frequently requested explicit rhythmic aids. To address timing guidance, we outline targeted visual and rhythmic additions in Limitations and Future Work (Section~\ref{sec:limitations}).

A notable diagnostic pattern in the \textit{Static} condition helps explain part of the pitch gap: several participants (e.g., \textbf{P1}, \textbf{P5}, \textbf{P15}) produced consistent \emph{octave-shift} errors (playing one octave too high) in retention tests. As shown in Table~\ref{tab:hand_similarity}, \textbf{P1}'s similarity score dropped sharply from the Static Immediate to the Static Retention phase, reflecting the impact of the octave transposition on positional alignment. This kind of systematic transposition inflates pitch error even when relative finger sequences are otherwise preserved. By contrast, \textit{Dynamic} guidance appears to stabilize spatial anchoring on the keyboard and to reinforce the intended fingering trajectory; this is consistent with its higher finger accuracy and lower error rates. The adaptively transparent exemplar reduces external guidance during stable passages and restores it after slips, providing just-in-time spatial landmarks (where to land, which finger to use) and temporal hints (when to move). Together, these help prevent drift into a transposed register and facilitate reinstatement of the learned pattern, thereby increasing retention.

\subsection{Hypotheses Revisited}

Based on the results, we return to our hypotheses.

\textbf{H1 (Skill transfer).} Evidence supports H1. As shown in the individual differences in Section~\ref{subsec:learning-curve}, at the switch to the second block (new melody and condition), participants who trained under \textit{Dynamic} first began at a higher initial performance (mean of the first two loops). This advantage held after controlling for melody and was visible within each melody; it also persisted to the end of the block with a larger share of participants exceeding the 0.60 criterion. These patterns indicate a transfer advantage that carries over to the new task setting.

\textbf{H2 (Retention).} Largely supported. \textit{Dynamic} showed smaller Immediate $\rightarrow$ Retention declines, especially for pitch and fingering, while timing remained comparable. This implies that correctness benefits stem from visibility adaptation, whereas rhythm likely requires explicit temporal cues.

\textbf{H3 (Training convergence).} Partially supported. Group learning curves (Sec.~\ref{subsec:learning-curve}) showed overall improvement under both conditions and a small \textit{Dynamic} advantage, but no reliable Condition$\times$Loop interaction, implying similar average learning rates. Crucially, at the individual level, the per-participant slope analysis (Figure~\ref{fig:indiv_errorrate}(b)) reveals a statistically reliable shift toward steeper error-rate reductions under \textit{Dynamic}: the mean $\Delta$\,slope (Dynamic$-$Static) $< 0$ with a 95\% CI entirely below $0$. Together with the larger share of \textit{Dynamic}-first learners exceeding the 0.60 criterion by the end of the second block (Figure~\ref{fig:switch_initial_and_end}(b)), these results indicate a modest but consistent convergence advantage for \textit{Dynamic}. The absence of a group-level interaction likely reflects learner heterogeneity and non-linear/ceiling effects that mask a subtle individual-level difference in the aggregate.


\subsection{Qualitative Feedback}

We analyzed open-ended responses from the post-study questionnaire (26/30 participants provided at least one comment). Three themes emerged.

\paragraph{What participants liked.}
The most frequent positives were \textit{finger positioning clarity} and \textit{followability}. Participants highlighted that the ghost hand made finger placement easy to imitate and track (11 mentions; e.g., \emph{“The position of the fingers is easy to follow.”}—\textbf{P10}; \emph{“…useful for practicing both timing and finger positions.”}—\textbf{P8}). Eight comments praised the overall clarity/guidance (\emph{“clear and easy to follow”}—\textbf{P2}). A few explicitly mentioned help with memory/location (2) and rhythm/timing (2), and some noted that the dynamic transparency felt natural (3; e.g., \emph{“looking back I actually like the concept and it felt natural”}—\textbf{P7}) and fun/motivating (2).

\paragraph{What was confusing or difficult.}
\textbf{General issues.} The most common issue was \textit{hand-tracking on ring/pinky fingers} (6; e.g., \emph{“The ring finger didn’t function too well.”}—\textbf{P2}; \emph{“…didn’t register properly when played with ring- or pinky finger.”}—\textbf{P8}). Some reported \textit{accidental keypresses} / touching neighboring keys (3; \textbf{P1}, \textbf{P7}, \textbf{P16}), \textit{audio/rhythm confusion} when both sources were audible (1; \textbf{P9}), and \textit{physical fatigue} (1; \textbf{P1}).

\textbf{Condition-specific.} In the \textit{Static} condition, participants mentioned \textit{occlusion/visibility} of ghost fingers by their own hands (1; \textbf{P4}) and \textit{afterimages} (1; \textbf{P1}). By contrast, several comments favored \textit{Dynamic} for being more intuitive and easier to follow, with fewer visibility/occlusion issues and less conflicting cues (e.g., \textbf{P21}, \textbf{P11}, \textbf{P19}, \textbf{P24}); some also noted that visibility recovered when dynamic fading became too light (e.g., \textbf{P28}).

\paragraph{Improvements suggested.}
Participants asked for (i) \textit{better tracking for ring/pinky} (3; \textbf{P8}, \textbf{P10}), (ii) \textit{alternative views / camera options} to improve \textit{viewpoint/line-of-sight} and judging contact (4; \textbf{P4}, \textbf{P9}), (iii) \textit{haptic/physical keyboard support} (3; \textbf{P10}, \textbf{P16}), (iv) \textit{beat/metronome indicators} (1; \textbf{P7}), and (v) \textit{notation/labels on keys} (1; \textbf{P1}). Several of these directly target the issues above—tracking, viewpoint/line-of-sight for contact judgment (with occlusion largely mitigated by \textit{Dynamic}), rhythm guidance, and input robustness.

\subsection{Implications and Generalizability}
Our findings suggest that dynamically modulated transparency reduces occlusion and focuses attention on \emph{where} to land and \emph{when} to move in visuomotor tasks that share three properties: (i) clear spatial targets, (ii) temporal sequencing, and (iii) intermittent self-occlusion. This is consistent with motor-learning evidence that adaptive or faded augmented feedback mitigates guidance dependence and supports retention \cite{winstein1994effects,anderson2005support}. Potential application areas include \emph{sports posture and rhythm drills} with AR/VR coaching \cite{witte2025sports}, \emph{handwriting and calligraphy} with guided stroke trajectories \cite{wu2022calligraphy}, and \emph{upper-limb rehabilitation} where VR/AR adjuncts support functional gains \cite{laver2011virtual}. However, \emph{domain-specific validation is required} before drawing conclusions.

We propose the following design guidelines: (1) Fade with proficiency, recover after slips. Treat guidance as a scaffold, not a crutch: use an opacity controller that decreases guidance visibility as the learner’s performance stabilizes and increases visibility after detectable errors; (2) Keep guidance visible. Composite the instructor above the learner’s representation with adaptive transparency, rendered without depth buffering, thereby keeping guidance visible while preventing occlusion by the environment; (3) Structure practice and probe retention. Support controlled, repeatable practice with task-aligned feedback: provide phrase-level chunking and looped replays, and follow with no-guidance probes to assess retention. These guidelines operationalize adaptive transparency as a strategy that reduces visual occlusion and cue dependence while enabling repeatable, task-aligned practice. These guidelines cover tasks that make use of clear spatial targets and temporal sequencing; assessing their validity in other domains remains future work.

\subsection{Limitations and Future Work}\label{sec:limitations}

\textbf{Algorithm Personalization.} Our study employed a fixed weighting scheme (e.g., pitch 70\%, timing 20\%, fingering 10\%) to calculate the performance score that drives transparency. While this weighting is grounded in pedagogical principles prioritizing note accuracy for beginners, a "one-size-fits-all" approach may not be optimal for every learner. For instance, a user who masters pitch quickly but struggles with rhythm might benefit from a higher weight on timing accuracy. Future work should investigate dynamic and personalized weighting schemes. This could involve allowing the system to automatically adjust weights based on a learner's recurring error patterns, or even providing users direct control to customize the feedback's focus according to their personal learning goals.

\textbf{Tracking and input robustness.}
Participants noted ring and pinky finger tracking errors and occasional stray key presses. In our current prototype, key presses are triggered solely by virtual contact between the tracked hand and the virtual keys; there is no physical keyboard or force-sensitive dynamics. As a result, learners cannot rely on tactile cues to judge when a key has been fully depressed, and small tracking jitters can lead to missed or spurious key events. In addition, both the learner’s tracked hand and the tutor’s ghost hand use the platform’s generic hand model with fixed proportions that do not exactly match each user’s hand size and span, which reduces ergonomic realism. Promising directions for future work include improving distal-joint tracking (e.g., with an auxiliary optical hand-tracking camera), adding light \textit{debounce}/\textit{false-touch} filtering, and personalizing hand scaling via a brief calibration step. Another avenue is to deploy the same adaptive transparency controller on top of a physical keyboard and real hands in AR/MR, where tactile feedback and visual appearance more closely match real-world piano study.

\textbf{View control and rhythm support.}
The current system offers a single first-person view and minimal tempo cues. Because timing differences between conditions were not reliable and participants suggested adding view options and rhythm aids, provide flexible view presets to keep the active region and the instructor visible; useful presets include snap to keyboard center, follow the instructor, and fit the active phrase region. Add rhythm cues such as an auditory metronome with a subtle visual pulse or a tempo strip, a one bar count-in, gradual tempo ramps, and optional beat subdivisions to support timing and improve visibility as well as temporal guidance.

\textbf{Instructor motion provenance.}
In this study, the instructor trajectories were recorded by the authors for proof of concept. Such recordings may encode idiosyncratic fingering and micro-timing that differ from expert pedagogy and across styles. Future work will capture demonstrations from professional piano instructors, compare expert- vs.\ self-recorded guidance, and explore aggregating or synthesizing instructor motions to provide style-appropriate cues.

\textbf{Generalizability and longevity.}
Our lab study focused on right-hand, monophonic, short phrases ($\approx$12\,s) in a controlled setting with immediate and 10-minute retention tests. This simplification was intentional to isolate the effect of adaptive visibility and ensure reliable hand-pose and key-press sensing in a within-subjects design, but it limits ecological validity relative to real piano learning (bimanual coordination, polyphony, expressive timing, longer phrases). It could be extend to bimanual/polyphonic repertoire, longer excerpts, and variable tempi/dynamics in authentic practice settings. Because the controller is performance-contingent (fade when stable, restore after slips) rather than score-specific, it should transfer naturally to these scenarios. In such settings, issues arising from self-occlusion and visual clutter are likely to be more pronounced, and we expect that our dynamic method would mitigate these effects more effectively than a static condition.


\section{Conclusion}
We introduced a skill-adaptive, dynamically transparent ghost instructor for VR piano training that scales its visibility to the learner’s moment-to-moment performance. The controller aggregates pitch, timing, and fingering accuracy into a weighted performance signal. This is mapped to instructor opacity with an asymmetric exponential moving average combined with a clamped visibility range, aiming to reduce cue dependence, and improve learning and short-term retention.

Across a within-subjects study, the \textit{Dynamic} mode consistently outperformed a \textit{Static} baseline. Participants achieved higher pitch and fingering accuracy and lower error rates under guidance, and these advantages persisted when cues were removed. Dynamic also limited performance decline from immediate to retention, while timing showed no reliable condition effect. Learning-curve analyses indicated steady improvement for both modes with a small but reliable advantage for \textit{Dynamic}. Subjectively, workload was lower and preferences favored \textit{Dynamic}.

Overall, \textit{Skill-Adaptive Ghost Instructors} helps learners internalize finger placement and motion patterns without fostering over-reliance on on-screen cues. Practically, we recommend fading assistance after correct behavior and reintroducing it after slips, capping visibility to preserve scene readability, and weighting feedback toward task-critical dimensions while monitoring technique. Beyond piano, dynamically transparent guidance is promising for skills combining spatial accuracy, rhythmic control, and ergonomics, including other instruments, upper-limb rehabilitation, and sports technique.

\section{Use of Large Language Models}
We used large language models solely for English language editing (grammar and wording). All study designs, analyses, and claims are the authors’ own, and no participant data were uploaded to any LLM services.

\begin{acks}
Research was supported by the e-DIPLOMA project (101061424)\footnote{\url{https://e-diplomaproject.eu/}}, funded by the European Union. Views and opinions expressed are, however, those of the authors only and do not necessarily reflect those of the European Union or the European Research Executive Agency (REA). Neither the European Union nor the granting authority can be held responsible for them.
\end{acks}

\bibliographystyle{ACM-Reference-Format}
\bibliography{references}










\end{document}